\documentclass[aps,prl,twocolumn,superscriptaddress, floatfix]{revtex4-2}  
\usepackage{dcolumn}
\usepackage{graphicx}
\usepackage{xcolor}
\usepackage{amssymb}
\usepackage{amsmath}
\usepackage{bm}
\usepackage{bbm}
\usepackage[colorlinks=true,linktocpage=true,breaklinks=true]{hyperref}
\usepackage{multirow}
\usepackage{braket}
\usepackage{soul}
\usepackage{appendix}
\usepackage{mathtools}
\usepackage{csquotes}
\usepackage[english]{babel}
\usepackage{subcaption}
\usepackage{mathrsfs}
\usepackage{subfiles}
\usepackage{booktabs}
\usepackage{tabularx}
\usepackage{float}
\usepackage{lipsum}
\usepackage{pgfplots}
\usepackage{mathdots}
\usepackage{tikz}
\usetikzlibrary{decorations.pathmorphing}
\usetikzlibrary{arrows.meta}
\usetikzlibrary{calc}
\usepgfplotslibrary{fillbetween}

\usepackage[nice]{nicefrac} 
\usepackage{braket} 
\usepackage{bbm} 
\usepackage{esint} 
\usepackage{cancel} 
\usepackage{mathtools} 
\usepackage{slashed} 
\usepackage{xargs} 

\let\oldepsilon\epsilon
\let\epsilon\varepsilon
\let\varepsilon\oldepsilon

\newcommand{\al}[1]{\begin{align} #1 \end{align}} 

\newcommand{\Exp}[1]{\text{e}^{#1}}		

\newcommand{\p}[1]{\left( #1 \right)}	
\newcommand{\cro}[1]{\left[ #1 \right]}	
\newcommand{\avg}[1]{\left\langle #1 \right\rangle} 
\newcommand{\acc}[1]{\left\lbrace #1 \right\rbrace} 

\newcommand{\ve}[1]{\mathbf{#1}} 

\newcommandx{\Sin}[2][1={}]{\text{sin}^{#1}\!\p{#2}}
\newcommandx{\Cos}[2][1={}]{\text{cos}^{#1}\!\p{#2}}
\newcommandx{\Tan}[2][1={}]{\text{tan}^{#1}\!\p{#2}}
\newcommandx{\Csc}[2][1={}]{\text{csc}^{#1}\!\p{#2}}
\newcommandx{\Sec}[2][1={}]{\text{sec}^{#1}\!\p{#2}}
\newcommandx{\Cot}[2][1={}]{\text{cot}^{#1}\!\p{#2}}
\newcommandx{\Arcsin}[2][1={}]{\text{arcsin}^{#1}\!\p{#2}}
\newcommandx{\Arccos}[2][1={}]{\text{arccos}^{#1}\!\p{#2}}
\newcommandx{\Arctan}[2][1={}]{\text{arctan}^{#1}\!\p{#2}}
\newcommandx{\Sinh}[2][1={}]{\text{sinh}^{#1}\!\p{#2}}
\newcommandx{\Cosh}[2][1={}]{\text{cosh}^{#1}\!\p{#2}}
\newcommandx{\Tanh}[2][1={}]{\text{tanh}^{#1}\!\p{#2}}

\newcommand{\Tr}[1]{\text{Tr}\p{#1}} 

\newcommand{\D}{\text{d}} 
\newcommandx{\dd}[3][1={},3={}]{\frac{\D^{#3}#1}{\D{#2}^{#3}}} 
\newcommand{\del}{\partial} 
\newcommandx{\ddp}[3][1={},3={}]{\frac{\del^{#3}#1}{\del{#2}^{#3}}} 

\newcommandx{\Int}[2][1={},2={}]{\int\displaylimits_{#1}^{#2}} 

\renewcommand{\Im}[1]{\text{Im}\acc{#1}} 

\renewcommand{\r}{\mathbf{r}} 

\newcommand{\op}[1]{\hat{\mathrm{#1}}}

\newcommand{\ketbra}[2]{\ket{#1}\!\!\bra{#2}}

\definecolor{COLOR1}{HTML}{dd4027} 
\definecolor{COLOR2}{HTML}{1c4286} 
\definecolor{COLOR3}{HTML}{f3a257} 

\begin{document}

\title{Symmetry-Resolved Spectral Weight Reveals Charge Localization in Correlated Insulator}

\author{Théo N. Dionne}
\email{theo.nathaniel.dionne@usherbrooke.ca}
\affiliation{Département de Physique et Institut Quantique, Université de Sherbrooke, Sherbrooke, J1K 2R1 Québec, Canada}
\affiliation{Regroupement Québécois sur les Matériaux de Pointe (RQMP), Québec, Canada}

\author{Maia G. Vergniory}
\affiliation{Département de Physique et Institut Quantique, Université de Sherbrooke, Sherbrooke, J1K 2R1 Québec, Canada}
\affiliation{Donostia International Physics Center (DIPC), Donostia-San Sebastián, Spain}
\affiliation{Regroupement Québécois sur les Matériaux de Pointe (RQMP), Québec, Canada}

\date{\today}


\begin{abstract}%
Can crystalline symmetry reveal where charge localizes in a strongly correlated insulator? We show that occupied symmetry-resolved spectral weights of the exact electronic Green's function define effective atomic limits. In an interacting Su-Schrieffer-Heeger chain, their displacement tracks the spin-resolved many-body polarization across distinct insulating phases. A quantum metric of the one-body reduced density matrix provides a complementary geometric characterization. Our results extend real-space symmetry analysis beyond occupied-band projectors.
\end{abstract}

\maketitle


\textit{Introduction}---The spatial organization of electronic charge provides a fundamental characterization of insulating phases. In noninteracting crystals, electric polarization, Wannier centers, and Topological Quantum Chemistry (TQC) connect momentum-space symmetry representations to real-space charge localization \cite{Bradlyn2017TQC,Po2017,Song2018,Kruthoff2017}. In particular, crystalline symmetries can identify atomic limits and obstructed insulating phases from the representations of occupied Bloch states. Extending this picture to strongly correlated systems is challenging, since the many-body ground state can no longer be described by a set of fully occupied single-particle bands. Although Green's function approaches and interacting extensions of band topology provide important insights \cite{Volovik2018,Iraola2021,Lessnich2021,soldini_interacting_2023,HerzogArbeitman2024,monsen2024supercellwannierfunctionsemergent}, establishing a direct connection between interacting single-particle information and real-space charge localization remains an outstanding challenge.

Recent developments have highlighted the potential of single-particle density matrices and quantum geometry for characterizing correlated electronic states. Natural orbitals of the interacting one-body reduced density matrix (1RDM) have been used to investigate topological phase transitions \cite{Zheng2018}, while other approaches construct cell natural orbitals starting from noninteracting band projectors to characterize the orbital structure relevant to interacting topological bands \cite{Verma2026}. Related studies have explored quantum geometry in fractionalized Mott phases \cite{Tang2026} and the information about many-body correlations encoded in fermionic covariance matrices \cite{Tarabunga2026}. These developments demonstrate the usefulness of one-particle descriptions, but also highlight the need to distinguish orbital symmetry labels from their interaction-dependent occupations. Starting from the electronic Green's function, we ask whether symmetry-resolved occupied spectral weights can be used to identify effective real-space charge centers and connect them to an independently calculated many-body observable.

In this Letter, we address this question by applying a symmetry-forward framework \cite{dionne2026characterizingmottinsulatorsinteracting} to an analytically solvable interacting extension \cite{mohamadi_emergence_2025,zhao_failure_2023} of the Su-Schrieffer-Heeger model \cite{OG_SSH,OG_SSH_2}. Crucially, our starting point is the exact single-particle Green's function of the interacting many-body system, rather than an assumed occupied-band projector. This incorporates the effects of electronic correlations on the single-particle excitation spectrum and spectral weights from the outset. Using Hatsugai-Kohmoto interactions \cite{OG_HK}, which provide analytical access to the many-body ground states and Green's function, we extract symmetry-resolved occupied spectral weights and construct effective atomic limits for three insulating phases. We demonstrate that the resulting charge-center displacements agree with the spin-resolved polarization calculated independently from the exact many-body ground states, even in the presence of fractional occupations and ground-state degeneracy. As a complementary characterization, we introduce the quantum volume of the interacting 1RDM to quantify the geometry explored in one-body state space. Our results thus establish a connection between fully interacting single-particle spectral information, real-space charge localization, and many-body polarization, providing a route toward symmetry-based characterization of correlated insulators using numerically accessible Green's functions \cite{senechal2004theoretical}.

\textit{SSH+HK Model}---We consider an interacting Su-Schrieffer-Heeger (SSH) chain \cite{OG_SSH,OG_SSH_2} with analytically tractable Hatsugai-Kohmoto (HK) interactions \cite{OG_HK,zhao_failure_2023,mohamadi_emergence_2025}. The model consists of two sublattices, $A$ and $B$, connected by intracell and intercell hoppings $v$ and $w$, respectively (Fig. \ref{fig:model}). A momentum-local interaction of strength $U$ allows the many-body ground states and single-particle Green's function to be obtained analytically. We fix the unit cell and inversion center as in Fig. \ref{fig:model} and use space group Pmmm, obtained by embedding the chain in three dimensions (see Supplemental Material IIB).

\begingroup
\captionsetup[subfigure]{
    position=top,
    labelformat=parens,
    labelsep=none,
    justification=raggedright,
    singlelinecheck=false,
    font={normalsize, bf},
    skip=0pt
}
\begin{figure}[htb!]
    \centering
    \begin{subfigure}[t!]{\linewidth}
        \centering
        \caption{}\label{fig:model}
        \resizebox{\linewidth}{!}{

\begin{tikzpicture}


\def \a {1}; 
\def \r {0.2}; 
\def \h {0.25}; 
\def \wiggleamplitude {0.7pt}; 
\def \wiggleperiod {4pt}; 
\def \onsiteloopradius {0.2\a}; 
\def \sitecontoursize {0.1pt}; 
\def \measurebarheight {-0.3\a}; 


\fill[COLOR3!20!white] (3.5,-0.65) rectangle (5.5, 0.65);


\foreach \i in {0,2,4}{
    \draw[very thick, COLOR1] (\a * \i,0) -- (\a * \i + 1,0);
}

\foreach \i in {1,3}{
    \draw[dashed, very thick, COLOR1] (\a * \i,0) -- (\a * \i + 1,0);
}
\draw[very thick, COLOR1] (-0.5,0) -- (0,0);
\draw[dashed, very thick, COLOR1] (5,0) -- (5.5,0);


\draw[very thick, decorate, decoration={snake, amplitude=\wiggleamplitude, segment length=\wiggleperiod}, COLOR3] 
    (0,0)
    arc[start angle=-90, end angle=360-90, radius=\onsiteloopradius]
;

\draw[very thick, decorate, decoration={snake, amplitude=\wiggleamplitude, segment length=\wiggleperiod}, COLOR3] 
    (0,0)
    arc[start angle=-180+60, end angle=-60, radius=2]
;

\draw[very thick, decorate, decoration={snake, amplitude=\wiggleamplitude, segment length=\wiggleperiod}, COLOR3] 
    (0,0)
    arc[start angle=-180+60, end angle=-60, radius=1]
;

\draw[very thick, decorate, decoration={snake, amplitude=\wiggleamplitude, segment length=\wiggleperiod}, COLOR3] 
    (0,0)
    arc[start angle=-180+60, end angle=-60, radius=3]
;

\draw[very thick, decorate, decoration={snake, amplitude=\wiggleamplitude, segment length=\wiggleperiod}, COLOR3] 
    (0,0)
    arc[start angle=-180+60, end angle=-89.5, radius=4]
;

\draw[very thick, decorate, decoration={snake, amplitude=\wiggleamplitude, segment length=\wiggleperiod}, COLOR3] 
    (0,0)
    arc[start angle=-60, end angle=-70, radius=3]
;


\foreach \i in {0,2,4}{
    \draw [black, line width=\sitecontoursize, fill=COLOR2] (\a * \i,0) circle (\r);
}

\foreach \i in {1,3,5}{
    \draw [black, line width=\sitecontoursize, fill=COLOR2] (\a * \i,0) circle (\r);
}

\node at (2*\a,0) {\raisebox{0.22ex}{\color{white}\textbf{A}}}; 
\node at (3*\a,0) {\hspace{0.1em}\color{white}\textbf{B}}; 

\node at (2.5 * \a,\h) {\large$\color{COLOR1}{v}$}; 
\node at (1.5 * \a,\h) {\large$\color{COLOR1}{w}$}; 
\node at (0.5,\h+0.2) {\color{COLOR3}\large${U}$}; 




\node at (4.5,0) {\color{COLOR2}$\boldsymbol{\times}$};
\node at (4.5,-0.3) {\color{COLOR2}$1a$};

\node at (5.5,0) {\color{COLOR2}$\boldsymbol{\times}$};
\node at (5.5,-0.3) {\color{COLOR2}$1b$};

\node at (4,0) {\color{white}$\boldsymbol{\times}$};
\node at (5,0) {\color{white}$\boldsymbol{\times}$};
\node at (5,0.35) {\color{COLOR2}$2i$};

\end{tikzpicture}
}
    \end{subfigure}
    \begin{subfigure}[t!]{\linewidth}
        \centering
        \caption{}\label{fig:GS}
        \begin{tikzpicture}[every node/.append style={scale=0.52}]
\pgfplotsset{compat=1.17}

\def\v{1}
\def\w{0.3}
\def\wiggleamplitude{0.6pt}
\def\wiggleperiod{3pt}
\pgfmathsetmacro{\DeltaLeft}{\v + \w}
\pgfmathsetmacro{\DeltaRight}{abs(\v - \w)}

\pgfplotsset{
    spectral axis/.style={
        width=0.31\linewidth,
        height=0.245\textwidth,
        scale only axis,
        xmin=-pi,
        xmax=pi,
        ymin=-\v-\w-0.3,
        ymax=\v+\w+0.3,
        xtick=\empty,
        ytick=\empty,
        axis line style={line width=1.2pt},
        label style={font=\large},
        clip=false,
    },
}

\begin{axis}[
    spectral axis,
    name=BIpU,
    ylabel={\LARGE$E$},
    ylabel style={rotate=-90},
]
\draw[COLOR3, dash pattern=on 3pt off 2pt, line width=1pt]
    (-pi,-\DeltaLeft) -- (pi,-\DeltaLeft);
\draw[COLOR3, dash pattern=on 3pt off 2pt, line width=1pt]
    (-pi,-\DeltaRight-\w) -- (pi,-\DeltaRight-\w);

\addplot[domain=-pi:pi,samples=100,color=COLOR1,line width=2pt]
    {sqrt(\v^2+\w^2+2*\v*\w*cos(deg(x)))};
\addplot[domain=-pi:pi,samples=100,color=COLOR2,line width=2pt]
    {-sqrt(\v^2+\w^2+2*\v*\w*cos(deg(x)))};

\node at (0,-\DeltaLeft) {\LARGE$\boldsymbol{\uparrow\downarrow}$};
\node at (-pi+0.4,-\DeltaRight-0.025) {\LARGE$\boldsymbol{\uparrow\downarrow}$};
\node at (pi-0.4,-\DeltaRight-0.025) {\LARGE$\boldsymbol{\uparrow\downarrow}$};
\node at (-pi/2+0.15,-\DeltaRight/2-\DeltaLeft/2-0.1)
    {\LARGE$\boldsymbol{\uparrow\downarrow}$};
\node at (pi/2-0.15,-\DeltaRight/2-\DeltaLeft/2-0.1)
    {\LARGE$\boldsymbol{\uparrow\downarrow}$};

\node at (0,\DeltaRight+0.15)
    {\LARGE\sffamily\textbf{BI+U}};
\node at (0,0)
    {\LARGE$
    {\color{COLOR2}\op{c}^{\dagger}_{k,-,\uparrow}
    \op{c}^{\dagger}_{k,-,\downarrow}}\ket{0}$};
\end{axis}

\begin{axis}[
    spectral axis,
    name=QFMI,
    at={(BIpU.east)},
    anchor=west,
    xlabel={\LARGE$k$},
]
\draw[COLOR3, dash pattern=on 3pt off 2pt, line width=1pt]
    (-pi,-\DeltaLeft) -- (pi,-\DeltaLeft);
\draw[COLOR3, dash pattern=on 3pt off 2pt, line width=1pt]
    (-pi,-\DeltaRight+0.20) -- (pi,-\DeltaRight+0.20);

\addplot[domain=-pi:pi,samples=100,color=COLOR1,line width=2pt]
    {sqrt(\v^2+\w^2+2*\v*\w*cos(deg(x)))};
\addplot[domain=-pi:pi,samples=100,color=COLOR2,line width=2pt]
    {-sqrt(\v^2+\w^2+2*\v*\w*cos(deg(x)))};

\node at (0,-\DeltaLeft) {\LARGE$\boldsymbol{\updownarrow}$};
\node at (-pi+0.3,-\DeltaRight-0.025) {\LARGE$\boldsymbol{\updownarrow}$};
\node at (pi-0.3,-\DeltaRight-0.025) {\LARGE$\boldsymbol{\updownarrow}$};
\node at (-pi/2+0.15,-\DeltaRight/2-\DeltaLeft/2-0.1)
    {\LARGE$\boldsymbol{\updownarrow}$};
\node at (pi/2-0.15,-\DeltaRight/2-\DeltaLeft/2-0.1)
    {\LARGE$\boldsymbol{\updownarrow}$};

\node at (0,\DeltaRight+0.15)
    {\LARGE\sffamily\textbf{QFMI}};
\node at (0,0)
    {\LARGE$\begin{aligned}
    \big\{{\color{COLOR2}\op{c}^{\dagger}_{k,-,\uparrow}}\ket{0}, {\color{COLOR2}\op{c}^{\dagger}_{k,-,\downarrow}}\ket{0}\!\!\big\}
    \end{aligned}$};
\end{axis}

\begin{axis}[
    spectral axis,
    name=HFMI,
    at={(QFMI.east)},
    anchor=west,
]
\draw[COLOR3, dash pattern=on 3pt off 2pt, line width=1pt]
    (-pi,-\DeltaLeft) -- (pi,-\DeltaLeft);
\draw[COLOR3, dash pattern=on 3pt off 2pt, line width=1pt]
    (-pi,+\DeltaLeft+0.20) -- (pi,+\DeltaLeft+0.20);

\addplot[domain=-pi:pi,samples=100,color=COLOR1,line width=2pt]
    {sqrt(\v^2+\w^2+2*\v*\w*cos(deg(x)))};
\addplot[domain=-pi:pi,samples=100,color=COLOR2,line width=2pt]
    {-sqrt(\v^2+\w^2+2*\v*\w*cos(deg(x)))};

\node at (0,-\DeltaLeft) {\LARGE$\boldsymbol{\updownarrow}$};
\node at (-pi+0.3,-\DeltaRight-0.025) {\LARGE$\boldsymbol{\updownarrow}$};
\node at (pi-0.3,-\DeltaRight-0.025) {\LARGE$\boldsymbol{\updownarrow}$};
\node at (-pi/2+0.15,-\DeltaRight/2-\DeltaLeft/2-0.1)
    {\LARGE$\boldsymbol{\updownarrow}$};
\node at (pi/2-0.15,-\DeltaRight/2-\DeltaLeft/2-0.1)
    {\LARGE$\boldsymbol{\updownarrow}$};

    \node at (0,\DeltaLeft) {\LARGE$\boldsymbol{\updownarrow}$};
    \node at (-pi+0.3,\DeltaRight+0.025) {\LARGE$\boldsymbol{\updownarrow}$};
    \node at (pi-0.3,\DeltaRight+0.025) {\LARGE$\boldsymbol{\updownarrow}$};
\node at (-pi/2+0.15,\DeltaRight/2+\DeltaLeft/2+0.1)
    {\LARGE$\boldsymbol{\updownarrow}$};
\node at (pi/2-0.15,\DeltaRight/2+\DeltaLeft/2+0.1)
    {\LARGE$\boldsymbol{\updownarrow}$};

\node at (0,\DeltaRight+0.15)
    {\LARGE\sffamily\textbf{HFMI}};
\node at (0,0)
    {\LARGE$\begin{aligned}
    \big\{&
    {\color{COLOR1}\op{c}^{\dagger}_{k,+,\uparrow}}
    {\color{COLOR2}\op{c}^{\dagger}_{k,-,\uparrow}}\ket{0},\\[-1mm]
    &{\color{COLOR1}\op{c}^{\dagger}_{k,+,\downarrow}}
    {\color{COLOR2}\op{c}^{\dagger}_{k,-,\downarrow}}\ket{0}
    \!\!\big\}
    \end{aligned}$};
\end{axis}

\end{tikzpicture}
    \end{subfigure}
    \caption{
        SSH+HK model. (a) Sites (blue) are connected by full/dashed lines (red) representing intra/inter-cell hoppings. HK interactions are indicated by wavy lines (orange) and relevant Wyckoff positions of Pmmm are marked by crosses in the unit cell (light rectangle). (b) Schematic of insulating ground states, where occupation of noninteracting bands is indicated by spin arrows and dashed orange lines set the interaction scale.
    }
\end{figure}
\endgroup

The corresponding lattice Hamiltonian, as constructed in Supplemental Material IA,
\al{
    \op{H} = \sum_{k,a,\sigma}\epsilon_{k,a}\op{n}_{k,a,\sigma}
    + U\sum_{k}\op{n}_{k,\uparrow}\op{n}_{k,\downarrow}\,,\label{eq:Hamiltonian}
}
is fully momentum-local due to the nature of HK interactions \cite{mohamadi_emergence_2025, zhao_failure_2023}. Here, $a=\pm$ labels the SSH bands, $\sigma$ the spin, and $k$ the crystal momentum in the first Brillouin zone (1BZ). The operator $\op{n}_{k,a,\sigma}$ denotes the corresponding occupation number, with missing band indices implicitly summed over. Particle-number conservation in each spin sector allows the symmetry and polarization analyses to be performed independently for each $\sigma$.

\textit{Insulating Phases}---We consider three distinct insulating phases of the SSH+HK chain, characterized by the interaction strength $U$ and electronic filling $\nu$ \cite{mohamadi_emergence_2025,zhao_failure_2023}. As illustrated in Fig. \ref{fig:GS}, these are (see Supplemental Material II):
\begin{enumerate}
    \item Interacting Band Insulator (\textbf{BI+U}),
    $\nu=1/2$ and $U<\text{gap}$;
    \item Quarter-Filled Mott Insulator (\textbf{QFMI}),
    $\nu=1/4$ and $U>\text{bandwidth}$;
    \item Half-Filled Mott Insulator (\textbf{HFMI}),
    $\nu=1/2$ and $U>\text{total bandwidth}$.
\end{enumerate}

While BI+U retains the ground state of the noninteracting SSH model, QFMI and HFMI exhibit highly degenerate many-body ground-state manifolds. Importantly, the two Mott phases differ in their orbital content: QFMI involves only the lower SSH band, whereas HFMI contains equal contributions from both bands. These three phases therefore provide a controlled setting to investigate whether symmetry-resolved single-particle information can distinguish different forms of correlated charge localization.

\textit{Interacting Single-Particle Picture}---At zero temperature, single-particle information on many-body phases is extracted via the single-particle Green's function of complex frequency $z$ \cite{rickayzen1980green, PyQCM}
\al{
    G_{\mu\nu}(z) &= \avg{
    \op{c}_{\mu}\frac{\mathbbm{1}}{z - \op{H} + E_\Omega}
    \op{c}_{\nu}^\dagger
    + \op{c}_{\nu}^\dagger\frac{\mathbbm{1}}{z + \op{H} - E_\Omega}
    \op{c}_{\mu}}\,.
}
In this section, Greek composite indices are defined as $\mu = (k,a,\sigma)$ and $\op{c}^{(\dagger)}_\mu$ are the corresponding electronic ladder operators. Moreover, averages are taken with respect to the generally degenerate ground-state manifold $\acc{\ket{\Omega_d}}$ of degeneracy $g$ and energy $E_\Omega$ as
\al{
    \avg{\circ} = \frac{1}{g}\sum_{d=1}^{g}\bra{\Omega_d}\circ\ket{\Omega_d}\,.\label{eq:averaging}
}
Since $G_{\mu\nu}(z)$ encodes single-particle excitation dynamics, it serves as an interacting counterpart to the band structure, with the irreducible representations (irreps) of the space group remaining meaningful labels in the presence of interactions \cite{dionne2026characterizingmottinsulatorsinteracting, soldini_interacting_2023}. In particular, contributions to the spectral function $A(\omega) = \Tr{-2\Im{\mathrm{G}(\omega + i0^+)}}$ are labelled by these irreps. In contrast to the noninteracting case, the spectral weight need not form sharp \enquote{bands} and its associated irrep content can be distributed over frequency \cite{dionne2026characterizingmottinsulatorsinteracting}.

For the SSH+HK chain (space group Pmmm), the isotropic orbitals at Wyckoff position 2i transform as the trivial representation $\mathrm{A}_g$ of group mmm. Inducing this representation to the full space group produces a band representation which is identified by the irreps of the space group at high-symmetry points $\Gamma=0$ and $\mathrm{X}=\pi/2a$ along the crystalline axis \cite{GraphTheoryMaia, Bradlyn2017TQC} (see Supplemental Material IB):
\al{
    \p{\mathrm{A}_g \uparrow G}_{2i}: \acc{\Gamma_1^{+}\oplus\Gamma_3^{-}, \mathrm{X}_1^{+}\oplus\mathrm{X}_3^{-}}\,.
}

The \textit{occupied} spectral weight (Fig. \ref{fig:spectrals}) of BI+U is identical to that of the noninteracting model, whereas that of QFMI is half-weighted due to quarter-filling. Conversely, the occupied spectral weight of HFMI is identical to a half-weighted equivalent of the noninteracting model's \textit{full} band structure.

\begingroup
\captionsetup[subfigure]{
    position=top,
    labelformat=parens,
    labelsep=none,
    justification=raggedright,
    singlelinecheck=false,
    font={normalsize, bf},
    skip=0pt
}
\begin{figure}[htb!]
    \centering
    \begin{subfigure}[t!]{0.65\linewidth}
        \centering
        \captionsetup{
            margin={6.6mm,0mm},
            skip=.4mm
        }
        \caption{}\label{fig:spectrals}
        \resizebox{\linewidth}{!}{\begin{tikzpicture}
    \pgfplotsset{
        compat=1.17,
        set layers,
    }
    \pgfplotsset{
        spectral axis/.style={
            width=0.6*9cm,
            height=0.6*7cm,
            scale only axis,
        },
    }

    \def\v{1}
    \def\w{0.5}

    \pgfmathsetmacro{\DeltaLeft}{\v + \w}
    \pgfmathsetmacro{\DeltaRight}{abs(\v - \w)}


    \begin{axis}[
        spectral axis,
        name=first plot,
        xmin=0,
        xmax=pi,
        xtick=\empty,
        ytick=\empty,
        xticklabels={},
        ylabel={{$\omega$}},
        ylabel style={rotate=-90},
        label style={font=\LARGE},
        axis line style={line width=1.5pt},
        clip=false,
    ]
        \begin{pgfonlayer}{axis background}
            \fill[COLOR3!20]
                (rel axis cs:0,0)
                rectangle
                (axis cs:pi,0);
        \end{pgfonlayer}

        \addplot[
            domain=0:pi,
            samples=100,
            color=COLOR1,
            line width=2pt,
            axis equal,
        ]
        {sqrt(\v^2 + \w^2 + 2*\v*\w*cos(deg(x)))};

        \addplot[
            domain=0:pi,
            samples=100,
            color=COLOR2,
            line width=2pt,
        ]
        {-sqrt(\v^2 + \w^2 + 2*\v*\w*cos(deg(x)))};

        \draw[COLOR1, fill]
            (0,\DeltaLeft) circle (2pt)
            node[left] {\Large$\Gamma_1^+$};

        \draw[COLOR2, fill]
            (0,-\DeltaLeft) circle (2pt)
            node[left] {\Large$\Gamma_3^-$};

        \draw[COLOR1, fill]
            (pi,\DeltaRight) circle (2pt)
            node[right] {\Large$\mathrm{X}_1^+[\mathrm{X}_3^-]$};

        \draw[COLOR2, fill]
            (pi,-\DeltaRight) circle (2pt)
            node[right] {\Large$\mathrm{X}_3^-[\mathrm{X}_1^+]$};

        \draw[dashed, line width=1pt]
            (0,0) -- (pi,0)
            node[near start, below] {\Large$\mu$};
    \end{axis}


    \begin{axis}[
        spectral axis,
        name=second plot,
        at=(first plot.south),
        anchor=north,
        xmin=0,
        xmax=pi,
        xtick=\empty,
        ytick=\empty,
        tick label style={font=\Large},
        axis line style={line width=1.5pt},
        clip=false,
    ]
        \begin{pgfonlayer}{axis background}
            \fill[COLOR3!20]
                (rel axis cs:0,0)
                rectangle
                (axis cs:pi,-0.25);
        \end{pgfonlayer}

        \addplot[
            domain=0:pi,
            samples=100,
            color=COLOR1,
            line width=2pt,
        ]
        {sqrt(\v^2 + \w^2 + 2*\v*\w*cos(deg(x)))};

        \addplot[
            domain=0:pi,
            samples=100,
            color=COLOR2,
            line width=2pt,
        ]
        {-sqrt(\v^2 + \w^2 + 2*\v*\w*cos(deg(x)))};

        \addplot[
            domain=0:pi,
            samples=100,
            color=COLOR1,
            line width=2pt,
        ]
        {sqrt(\v^2 + \w^2 + 2*\v*\w*cos(deg(x))) + 2*\w + 0.5};

        \addplot[
            domain=0:pi,
            samples=100,
            color=COLOR2,
            line width=2pt,
        ]
        {-sqrt(\v^2 + \w^2 + 2*\v*\w*cos(deg(x))) + 2*\w + 0.5};

        \draw[COLOR1, fill]
            (0,\DeltaLeft) circle (2pt)
            node[left] {\Large$\frac{1}{2}\Gamma_1^+$};

        \draw[COLOR1, fill]
            (0,\DeltaLeft + 2*\w + 0.5) circle (2pt)
            node[left] {\Large$\frac{1}{2}\Gamma_1^+$};

        \draw[COLOR2, fill]
            (0,-\DeltaLeft) circle (2pt)
            node[left] {\Large$\frac{1}{2}\Gamma_3^-$};

        \draw[COLOR2, fill]
            (0,-\DeltaLeft + 2*\w + 0.5) circle (2pt)
            node[left] {\Large$\frac{1}{2}\Gamma_3^-$};

        \draw[COLOR1, fill]
            (pi,\DeltaRight) circle (2pt)
            node[below right]
            {\Large$\frac{1}{2}\mathrm{X}_1^+[\mathrm{X}_3^-]$};

        \draw[COLOR1, fill]
            (pi,\DeltaRight + 2*\w + 0.5) circle (2pt)
            node[above right]
            {\Large$\frac{1}{2}\mathrm{X}_1^+[\mathrm{X}_3^-]$};

        \draw[COLOR2, fill]
            (pi,-\DeltaRight) circle (2pt)
            node[below right]
            {\Large$\frac{1}{2}\mathrm{X}_3^-[\mathrm{X}_1^+]$};

        \draw[COLOR2, fill]
            (pi,-\DeltaRight + 2*\w + 0.5) circle (2pt)
            node[above right]
            {\Large$\frac{1}{2}\mathrm{X}_3^-[\mathrm{X}_1^+]$};

        \draw[dashed, line width=1pt]
            (0,-0.25) -- (pi,-0.25)
            node[near start, below] {\Large$\mu$};
    \end{axis}


    \begin{axis}[
        spectral axis,
        name=third plot,
        at=(second plot.south),
        anchor=north,
        xmin=0,
        xmax=pi,
        xtick=\empty,
        ytick=\empty,
        xlabel={$k$},
        label style={font=\LARGE},
        axis line style={line width=1.5pt},
        clip=false,
    ]
        \begin{pgfonlayer}{axis background}
            \fill[COLOR3!20]
                (rel axis cs:0,0)
                rectangle
                (axis cs:pi,{2*\DeltaLeft});
        \end{pgfonlayer}

        \addplot[
            domain=0:pi,
            samples=100,
            color=COLOR1,
            line width=2pt,
        ]
        {sqrt(\v^2 + \w^2 + 2*\v*\w*cos(deg(x)))};

        \addplot[
            domain=0:pi,
            samples=100,
            color=COLOR2,
            line width=2pt,
        ]
        {-sqrt(\v^2 + \w^2 + 2*\v*\w*cos(deg(x)))};

        \addplot[
            domain=0:pi,
            samples=100,
            color=COLOR1,
            line width=2pt,
        ]
        {sqrt(\v^2 + \w^2 + 2*\v*\w*cos(deg(x))) + 4*\DeltaLeft};

        \addplot[
            domain=0:pi,
            samples=100,
            color=COLOR2,
            line width=2pt,
        ]
        {-sqrt(\v^2 + \w^2 + 2*\v*\w*cos(deg(x))) + 4*\DeltaLeft};

        \draw[COLOR1, fill]
            (0,\DeltaLeft) circle (2pt)
            node[left] {\Large$\frac{1}{2}\Gamma_1^+$};

        \draw[COLOR1, fill]
            (0,\DeltaLeft + 4*\DeltaLeft) circle (2pt)
            node[left] {\Large$\frac{1}{2}\Gamma_1^+$};

        \draw[COLOR2, fill]
            (0,-\DeltaLeft) circle (2pt)
            node[left] {\Large$\frac{1}{2}\Gamma_3^-$};

        \draw[COLOR2, fill]
            (0,-\DeltaLeft + 4*\DeltaLeft) circle (2pt)
            node[left] {\Large$\frac{1}{2}\Gamma_3^-$};

        \draw[COLOR1, fill]
            (pi,\DeltaRight) circle (2pt)
            node[above right]
            {\Large$\frac{1}{2}\mathrm{X}_1^+[\mathrm{X}_3^-]$};

        \draw[COLOR1, fill]
            (pi,\DeltaRight + 4*\DeltaLeft) circle (2pt)
            node[above right]
            {\Large$\frac{1}{2}\mathrm{X}_1^+[\mathrm{X}_3^-]$};

        \draw[COLOR2, fill]
            (pi,-\DeltaRight) circle (2pt)
            node[below right]
            {\Large$\frac{1}{2}\mathrm{X}_3^-[\mathrm{X}_1^+]$};

        \draw[COLOR2, fill]
            (pi,-\DeltaRight + 4*\DeltaLeft) circle (2pt)
            node[below right]
            {\Large$\frac{1}{2}\mathrm{X}_3^-[\mathrm{X}_1^+]$};

        \draw[dashed, line width=1pt]
            (0,2*\DeltaLeft) -- (pi,2*\DeltaLeft)
            node[near start, below] {\Large$\mu$};
    \end{axis}

    \node[
        at=(first plot.center),
        xshift=-0.2cm,
        yshift=0.6cm,
        text width=1cm,
    ] {\Large\sffamily\textbf{BI+U}};

    \node[
        at=(second plot.center),
        xshift=-0.2cm,
        yshift=0.85cm,
        text width=1cm,
    ] {\Large\sffamily\textbf{QFMI}};

    \node[
        at=(third plot.center),
        xshift=-0.2cm,
        yshift=1.18cm,
        text width=1cm,
    ] {\Large\sffamily\textbf{HFMI}};

\end{tikzpicture}}
    \end{subfigure}%
    \begin{subfigure}[t!]{0.35\linewidth}
        \centering
        \caption{}\label{fig:AL_BIpU_QFMI}
        \resizebox{\linewidth}{!}{

\begin{tikzpicture}

    \begin{scope}
        \fill[COLOR3!20] (0,-0.65) rectangle (2,0.65);
        \coordinate (T1a) at (1,0);
        \coordinate (T1b) at (2,0);

        \filldraw[COLOR2]
            ($(T1a)-(0.40cm,0)$) ellipse (0.4cm and 0.25cm)
            node[yshift=-0.40cm] {$\mathrm{B}_{3u}$};
        \filldraw[COLOR2]
            ($(T1a)+(0.40cm,0)$) ellipse (0.4cm and 0.25cm);

        \node at ($(T1a)-(0.6cm,0)$) {\color{white}$\bm{+}$};
        \node at ($(T1a)+(0.6cm,0)$) {\color{white}$\bm{-}$};

        \fill[COLOR3] (T1a) circle (1.5pt)
            node[xshift=-0.25cm] {\color{COLOR3}$1a$};
        \fill[black] (T1b) circle (1.5pt)
            node[xshift=0.25cm] {$1b$};

        \node at ($(T1a)+(0,0.45cm)$) {\large$w<v$};
    \end{scope}

    \begin{scope}[yshift=-1.35cm]
        \fill[COLOR3!20] (0,-0.65) rectangle (2,0.65);
        \coordinate (B1a) at (1,0);
        \coordinate (B1b) at (2,0);

        \filldraw[COLOR2]
            ($(B1b)-(0.40cm,0)$) ellipse (0.4cm and 0.25cm)
            node[yshift=-0.40cm] {$\mathrm{B}_{3u}$};
        \filldraw[COLOR2]
            ($(B1b)+(0.40cm,0)$) ellipse (0.4cm and 0.25cm);

        \node at ($(B1b)-(0.6cm,0)$) {\color{white}$\bm{+}$};
        \node at ($(B1b)+(0.6cm,0)$) {\color{white}$\bm{-}$};

        \fill[black] (B1a) circle (1.5pt)
            node[xshift=-0.25cm] {$1a$};
        \fill[COLOR3] (B1b) circle (1.5pt)
            node[xshift=0.25cm] {\color{COLOR3}$1b$};

        \node at ($(B1a)+(0,0.45cm)$) {\large$w>v$};
    \end{scope}
\end{tikzpicture}}
        \vspace{1cm}
        \caption{}\label{fig:AL_HFMI}
        \resizebox{\linewidth}{!}{

\begin{tikzpicture}

    \begin{scope}
        \fill[COLOR3!20] (0,-0.65) rectangle (2,0.65);
        \coordinate (T1a) at (1,0);
        \coordinate (T1b) at (2,0);

        \fill[COLOR1] (T1a) circle (0.4cm)
            node[yshift=0.45cm,xshift=0.41cm] {$\mathrm{A}_{1g}$};
        \filldraw[COLOR2]
            ($(T1a)-(0.40cm,0)$) ellipse (0.4cm and 0.25cm)
            node[yshift=-0.40cm] {$\mathrm{B}_{3u}$};
        \filldraw[COLOR2]
            ($(T1a)+(0.40cm,0)$) ellipse (0.4cm and 0.25cm);

        \node at ($(T1a)-(0.60cm,0)$) {\color{white}$\bm{+}$};
        \node at ($(T1a)+(0.60cm,0)$) {\color{white}$\bm{-}$};

        \fill[COLOR3] (T1a) circle (1.5pt)
            node[xshift=-0.25cm] {\color{COLOR3}$1a$};
        \fill[black] (T1b) circle (1.5pt)
            node[xshift=0.25cm] {$1b$};
    \end{scope}

    \begin{scope}[yshift=-1.35cm]
        \fill[COLOR3!20] (0,-0.65) rectangle (2,0.65);
        \coordinate (B1a) at (1,0);
        \coordinate (B1b) at (2,0);

        \fill[COLOR1] (B1b) circle (0.4cm)
            node[yshift=0.45cm,xshift=0.41cm] {$\mathrm{A}_{1g}$};
        \filldraw[COLOR2]
            ($(B1b)-(0.40cm,0)$) ellipse (0.4cm and 0.25cm)
            node[yshift=-0.40cm] {$\mathrm{B}_{3u}$};
        \filldraw[COLOR2]
            ($(B1b)+(0.40cm,0)$) ellipse (0.4cm and 0.25cm);

        \node at ($(B1b)-(0.60cm,0)$) {\color{white}$\bm{+}$};
        \node at ($(B1b)+(0.60cm,0)$) {\color{white}$\bm{-}$};

        \fill[black] (B1a) circle (1.5pt)
            node[xshift=-0.25cm] {$1a$};
        \fill[COLOR3] (B1b) circle (1.5pt)
            node[xshift=0.25cm] {\color{COLOR3}$1b$};
    \end{scope}

\end{tikzpicture}}
    \end{subfigure}
    ~
    \begin{subfigure}[t!]{0.45\linewidth}
        \centering

    \end{subfigure}
    \caption{
        Interacting single-particle analysis of the correlated insulating phases of the SSH+HK chain. Red and blue denote contributions inherited from the upper and lower noninteracting bands, respectively. (a) Spin-resolved spectral weight as a function of crystal momentum $k$. The shaded region lies below the chemical potential $\mu$; bracketed irreps denote the labels for $w>v$. (b) Effective atomic limits of BI+U and QFMI, exhibiting a displacement from $1a$ to $1b$ across $w=v$. (c) The simultaneous $1a$- and $1b$-centered effective atomic limits of HFMI, which remain unchanged across $w=v$.
    }
\end{figure}
\endgroup

Complementarily, the concept of a projector onto the subspace of occupied bands is generalized by the single-particle reduced density matrix (1RDM) which relates to average electron density \cite{dionne2026characterizingmottinsulatorsinteracting}. 1RDMs for each phase, as calculated from the Green's function \cite{PyQCM, dionne2026characterizingmottinsulatorsinteracting}:
\al{
    \gamma_{\mu\nu} = \avg{\op{c}^\dagger_\nu\op{c}_\mu} \overset{T=0}{=} \int_{\mathcal{C}_<}\frac{\D{z}}{2\pi i}G_{\mu\nu}(z)\,,
}
are tabulated in Table \ref{tab:1RDMs} where the basis is spanned by
\al{
    \ket{k,\pm,\sigma} = \frac{1}{\sqrt{2}}\cro{\Exp{i\phi(k)}\ket{k,A,\sigma} \pm \ket{k,B,\sigma}}\,,\label{eq:eigenstates}
}
with
\al{
    \phi(k) \equiv \arg\left(v+w e^{2iak}\right)\,.\label{eq:phase_angle}
}
The phase $\phi(k)$ is chosen continuously along the Brillouin zone away from the gap-closing point $w=v$.

\begin{table}[H]
\centering
\setlength{\tabcolsep}{15pt}
\renewcommand{\arraystretch}{1.2}
\begin{tabularx}{\columnwidth}{l l}
\toprule
Phase & 1RDM \\ \midrule
BI+U & $\ketbra{k,-,\sigma}{k,-,\sigma}$ \\
QFMI & $\frac{1}{2}\ketbra{k,-,\sigma}{k,-,\sigma}$ \\
HFMI & $\frac{1}{2}\ketbra{k,-,\sigma}{k,-,\sigma} + \frac{1}{2}\ketbra{k,+,\sigma}{k,+,\sigma}$ \\\bottomrule
\end{tabularx}
\caption{1RDMs of the SSH+HK model's insulators in the basis of noninteracting eigenstates.}
\label{tab:1RDMs}
\end{table}

Taken together, the symmetry-resolved spectral functions and 1RDMs form an interacting single-particle picture which provides a straightforward symmetry-based adaptation of band theory in the strongly correlated regime. Crucially, this single-particle description (Fig. \ref{fig:spectrals} and Table \ref{tab:1RDMs}) suggests an interpretation of the correlated insulators in the SSH+HK model in terms of their noninteracting counterpart \cite{OG_SSH, OG_SSH_2}: both BI+U and QFMI resemble the half-filled SSH chain while HFMI resembles the fully filled SSH chain.


\textit{Effective Atomic Limits}---The organization of occupied spectral weight in real space is characterized through effective atomic limits. Given the model \eqref{eq:Hamiltonian}, we perform the decomposition independently in each spin sector. Analogously to the noninteracting case, the relevant combinations of irreps are broken down into elementary band representations (EBRs) in reciprocal space which are in turn related to effective atomic limits in real space. All relevant EBRs and atomic limits are presented in Supplemental Material IV.

In conventional TQC, atomic limits are identified from integer multiplicities of occupied band representations. Here, we extend this construction to interaction-dependent symmetry-resolved spectral weights obtained from the exact single-particle Green's function, whose coefficients need not be integers. The QFMI provides a particularly instructive example: although its occupied states carry the same symmetry representations as those of BI+U, their spectral weight is reduced by a factor of two. Consequently, an analysis based on symmetry eigenstates alone cannot distinguish these phases, whereas the spectral-weight decomposition captures the corresponding reduction in their spin-resolved many-body polarization. By contrast, HFMI has equal contributions associated with two distinct atomic limits. Thus, the symmetry content and fractional weights of interacting single-particle states provide complementary information that is essential for characterizing charge localization beyond conventional band theory.

As can be observed in Figs. \ref{fig:AL_BIpU_QFMI} and \ref{fig:AL_HFMI}, the effective atomic limits draw a sharp distinction between HFMI and the remaining phases. Whereas HFMI admits the same two distinct effective atomic limits for \textit{any} $w$ (one at $1a$ and the other at $1b$), BI+U and QFMI possess a unique effective atomic limit centered at $1a$ when $w<v$ and $1b$ when $w>v$. This identifies BI+U and QFMI, but \textit{not} HFMI, as possessing a parameter-dependent shift in their effective charge centers.


\textit{Polarization}---Rigorous assessment of charge localization is carried out using the spin-resolved polarization \cite{mohamadi_emergence_2025, Budich2013} \al{
    P_{\sigma} = \lim_{L\to\infty}\frac{e}{2\pi}\Im{\ln\avg{\Exp{2\pi i\op{X}_\sigma/L}}}  \label{eq:Polarization}\,,
} where $\op{X}_\sigma \equiv \sum_{j}x_j\op{n}_{j,\sigma}$ (see Supplemental Material VA) and the average is taken in the ground-state manifold as \eqref{eq:averaging}.

\begin{figure}[htb!]
    \centering
    \includegraphics[width=\linewidth]{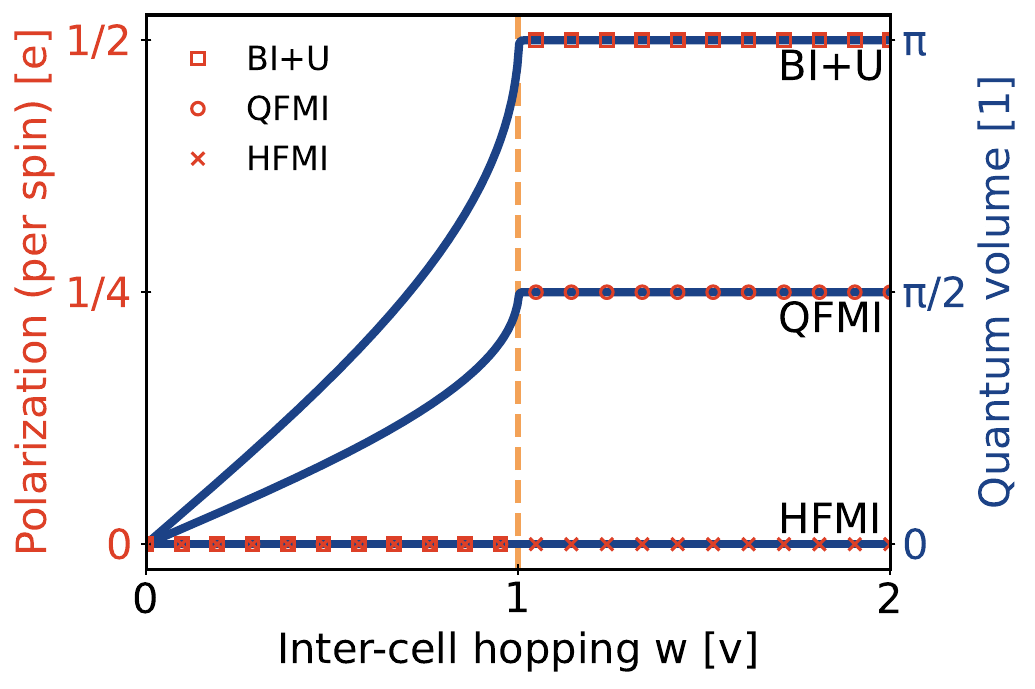}
    \caption{
        Spin-resolved polarization $P_\sigma$ (red points) and quantum volume $\mathscr{V}$ (blue curve) as a function of inter-cell hopping $w$. For BI+U and QFMI, the onset of nonzero polarization coincides with saturation of quantum volume at $w=v$, whereas both vanish for HFMI.
    }
    \label{fig:PolarizationAndVolume}
\end{figure}

As established from the Hamiltonian \eqref{eq:Hamiltonian}, Resta’s construction \cite{RestaCanonicalPol} can be applied independently to each spin sector. As calculated in Supplemental Material VB, this quantity distinguishes the insulating ground states of the SSH+HK chain (Fig. \ref{fig:PolarizationAndVolume}).

The spin-resolved polarization of BI+U shows a transition from $0$ to $e/2$ across $w=v$ while QFMI's response is halved and HFMI remains pinned at $0$ regardless of $w$. This behavior follows naturally as the ground state of BI+U is a copy of the spinless SSH chain's ground state in each spin channel. For QFMI, the degenerate ground-state manifold averaging allows the spin-resolved polarization to differ from the usual inversion-quantized values (see Supplemental Material VB). Evaluating this average yields half of BI+U's response, that is $e/4$ for $w>v$. Conversely, complete occupation of the lower and upper bands in each HFMI branch is unitarily equivalent to filling both localized sublattice orbitals, yielding a fixed atomic limit independent of $w$ (see Supplemental Material IIA).

Remarkably, the polarizations of BI+U and QFMI mirror their effective single-particle atomic limits, whereas HFMI shows no change in polarization across $w=v$. This agreement promotes the idea that EBR analysis of occupied spectral weight yields valuable information about charge even when electrons interact strongly.


\textit{Single-Particle Quantum Volume}---A complementary geometric characterization of the interacting insulators in the SSH+HK chain is provided by the quantum volume of their 1RDMs, which corresponds to the metric length of the 1RDM image in this one-dimensional model.

We employ a manifestly gauge-invariant quantum metric \cite{QGT_Mixed, QGT_PRB} defined directly from the 1RDM,
\al{
    g_{kk}^{(\text{1RDM})} = \frac{1}{2}\operatorname{Tr}\cro{\p{\mathcal{D}_k\ve{\gamma}}^\dagger\mathcal{D}_k\ve{\gamma}}\,.
}
Here, $\mathcal{D}_k\ve{\gamma}$ is the covariant derivative (as derived in Supplemental Material VI). This metric reduces to the standard quantum metric when the 1RDM is a projector, as in the band theory limit. Since the metric encodes distances on the space of states, it can be integrated to calculate the volume of the manifold of single-particle states over the 1BZ:
\al{
    \mathscr{V} = \int_{\text{1BZ}}\sqrt{g_{kk}^{(\text{1RDM})}}\D{k}\,,
}
where $k\in[-\pi/2a,\pi/2a]$ and $\mathscr{V}$ is evaluated separately in each spin sector (see Supplemental Material VI for a schematic illustration). As shown in Fig. \ref{fig:PolarizationAndVolume}, the quantum volume as a function of inter-cell hopping $w$ distinguishes between all three phases.

For BI+U and QFMI, increasing $w$ enlarges the range of $\phi(k)$ \eqref{eq:phase_angle} explored over the 1BZ and therefore the extent of the 1RDM image in state space. Upon entering the regime $w>v$, the image covers all of one-body state space and the quantum volume reaches its maximal value. The factor of two between the curves follows from $\gamma_{\text{QFMI}}=\frac{1}{2}\gamma_{\text{BI+U}}$. By contrast, $\gamma_{\text{HFMI}}=\frac{1}{2}\mathbbm{1}$ is independent of $k$ and $w$, so its image is a point and its quantum volume vanishes. This geometric characterization is complementary to the symmetry-based identification of effective atomic limits and does not establish a universal relation between quantum volume and many-body polarization.

\textit{Conclusion}---We have demonstrated how symmetry-resolved occupied spectral weights extracted from the exact interacting single-particle Green's function can be used to construct effective atomic limits, extending the symmetry-forward framework of Ref. \cite{dionne2026characterizingmottinsulatorsinteracting}. In the SSH+HK model, BI+U and QFMI share the same symmetry representations but differ in their fractional spectral weights, whereas HFMI contains contributions from two distinct atomic limits. These results show that symmetry labels alone are insufficient to characterize the effective real-space organization of charge when interactions produce fractional single-particle occupations.

The quantum volume of the 1RDM provides a complementary characterization of the one-body geometry, distinguishing momentum-dependent electronic textures from the geometrically trivial HFMI. This perspective complements recent developments connecting quantum geometry to topological and correlated systems \cite{mera_LowerBound_2021,mera2_LowerBound_2022, Bernevig_QG_2025,Queiroz_QG_2026}. Although the momentum-local nature of HK interactions makes the model analytically tractable, our construction relies on quantities accessible through interacting single-particle Green's functions, which opens a route toward extending symmetry-based descriptions of charge localization to more general correlated materials.


\textit{Acknowledgments}---We are grateful to André-Marie Tremblay for insightful discussions and helpful suggestions in the writing of this manuscript. We also thank Guangze Chen \cite{TopologicalZeroModesChen} and Carlos Mejuto-Zaera \cite{gGutTopoCarlos} for sharing their interesting work.

T.N.D. acknowledges the support of the Natural Sciences and Engineering Research Council of Canada (NSERC), the Fonds de recherche du Québec - Nature et technologies (FRQNT) and the Fondation de l'Université de Sherbrooke (FUS).

M.G.V. received financial support from the Canada Excellence Research Chairs Program for Topological Quantum Matter and PID2025-172166NB-C22 funded by MICIU/AEI/10.13039/501100011033


\bibliography{ref}

\end{document}


\title{Supplemental Material: Symmetry-Resolved Spectral Weight Reveals Charge Localization in Correlated
Insulator}

\author{Théo N. Dionne}
\email{theo.nathaniel.dionne@usherbrooke.ca}
\affiliation{Département de Physique et Institut Quantique, Université de Sherbrooke, Sherbrooke, J1K 2R1 Québec, Canada}
\affiliation{Regroupement Québécois sur les Matériaux de Pointe (RQMP), Québec, Canada}

\author{Maia G. Vergniory}
\affiliation{Département de Physique et Institut Quantique, Université de Sherbrooke, Sherbrooke, J1K 2R1 Québec, Canada}
\affiliation{Donostia International Physics Center (DIPC), Donostia-San Sebastián, Spain}
\affiliation{Regroupement Québécois sur les Matériaux de Pointe (RQMP), Québec, Canada}

\date{\today}


\setcounter{equation}{0}
\setcounter{figure}{0}
\setcounter{table}{0}

\renewcommand{\theequation}{S\arabic{equation}}
\renewcommand{\thefigure}{S\arabic{figure}}
\renewcommand{\thetable}{S\arabic{table}}

\maketitle

\tableofcontents

\section{SSH+HK model} \label{sec:SSHHK_model} 

\subsection{Construction} \label{SSH_model_construction} 

The SSH+HK model \cite{mohamadi_emergence_2025} is a Su-Schrieffer-Heeger chain \cite{OG_SSH, OG_SSH_2} with orbital-isotropic \cite{zhao_failure_2023} Hatsugai-Kohmoto interactions \cite{OG_HK}. Starting in real space,
\al{
    \op{H} &= -\sum_{i\sigma} \cro{v\op{c}_{i,B,\sigma}^{\dagger}\op{c}_{i,A,\sigma} + w\op{c}_{i+1,A,\sigma}^{\dagger}\op{c}_{i,B,\sigma}+ \text{h.c.}} + U\sum_{ij}\p{\op{n}_{i,A,\uparrow} + \op{n}_{i,B,\uparrow}}\p{\op{n}_{j,A,\downarrow} + \op{n}_{j,B,\downarrow}}\,,\label{eq:SSHHK_Hamiltonian_real_space}
}
where $v,w,U\in\mathbb{R}^+$, $\op{c}_{i,a,\sigma}^{(\dagger)}$ annihilates(creates) an electron at site $i$ in orbital $a$ with spin $\sigma$ and $
\op{n} = \op{c}^\dagger\op{c}$. It is important to note that these interactions do not change form when applying a unitary transformation in orbital space, since
\begin{equation}
	\begin{aligned}
    	\op{U}\op{N}_\sigma\op{U}^\dagger &= \sum_{i,a}\op{U}\op{c}_{i,a,\sigma}^\dagger\op{U}^\dagger\op{U}\op{c}_{i,a,\sigma}\op{U}^\dagger\\
        &= \sum_{i,a,b,b'}U_{a,b}^*\op{c}_{i,b,\sigma}^\dagger\op{c}_{i,b',\sigma}U_{b',a}\\
        &= \sum_{i,b}\op{c}_{i,b,\sigma}^\dagger\op{c}_{i,b,\sigma}\\
        &= \op{N}_\sigma\,.
	\end{aligned}
\end{equation}

Next, the following convention for Fourier transforms,
\begin{align}
    \op{c}_{k,s,\sigma} &= \frac{1}{\sqrt{N}}\sum_{j}\Exp{-ik(2a)j}\op{c}_{j,s,\sigma} &
    \op{c}_{j,s,\sigma} &= \frac{1}{\sqrt{N}}\sum_{k}\Exp{+ik(2a)j}\op{c}_{k,s,\sigma}\,,
\end{align}
is used to rewrite the Hamiltonian in reciprocal space:
\begin{align}
        \op{H} &= -\sum_{k\sigma} \cro{v\op{c}_{k,B,\sigma}^{\dagger}\op{c}_{k,A,\sigma} + w\Exp{-ik2a}\op{c}_{k,A,\sigma}^{\dagger}\op{c}_{k,B,\sigma}+ \text{h.c.}} + U\sum_{k}\p{\op{n}_{k,A,\uparrow} + \op{n}_{k,B,\uparrow}}\p{\op{n}_{k,A,\downarrow} + \op{n}_{k,B,\downarrow}}\,.
\end{align}
This k-local Hamiltonian
\al{
    \op{H} = \sum_{k,a,\sigma}\epsilon_{k,a,\sigma}\op{n}_{k,a,\sigma} + U\sum_{k}\op{n}_{k\uparrow}\op{n}_{k\downarrow}
}
can be rendered diagonal in the SSH band basis
\al{
    \op{c}^{\dagger}_{k,\pm,\sigma} = \frac{1}{\sqrt{2}}\cro{\Exp{i\phi(k)}\op{c}^{\dagger}_{k,A,\sigma} \pm \op{c}^{\dagger}_{k,B,\sigma}}
}
where
\al{
    \phi(k) = \arg\p{v+w e^{2iak}}
}
along with the associated eigenenergies of the noninteracting portion:
\al{
    \epsilon_{k,\pm,\sigma} = \pm\sqrt{v^2 + w^2 + 2vw\Cos{2ka}}.
}

\subsection{Space Group Symmetry} \label{SSH_model_symmetry} 

By embedding the one-dimensional chain into 3D, we may use space groups to characterize the symmetry of the model \cite{Iraola2021, dionne2026characterizingmottinsulatorsinteracting}. Upon imposing an s-like character on the orbitals at $A$ and $B$, the real-space Hamiltonian of the SSH+HK chain \eqref{eq:SSHHK_Hamiltonian_real_space} can be understood as having a Pmmm symmetry group. Whereas one mirror axis acts along the crystalline axis as $j\mapsto-j$ and $A\mapsto B$, both others act trivially. Crucially, we may treat each spin projection individually since the Hamiltonian is $\op{S}_z$ conserving.

Within the site-symmetry group of Pmmm, the spatially even orbitals are placed at Wyckoff position (WP) $2i$, hence transforming under the trivial irrep $A_1$. Using the Bilbao Crystallographic Server \cite{Bradlyn2017TQC, GraphTheoryMaia}, these orbitals induce
\al{
    \p{\mathrm{A}_g \uparrow G(2)}_{2i}: \acc{\Gamma_1^{+}\oplus\Gamma_3^{-}, \mathrm{X}_1^{+}\oplus\mathrm{X}_3^{-}}\,.
}
Given the $k$-local form of the model and the ground states of the interacting system (Sec. \ref{sec:GS}), the symmetry labels of the noninteracting bands carry over naturally to the interacting phases considered in this work.

Since each high-symmetry point possesses one even and one odd irreducible representation (irrep) of the space group, they can be easily assigned to the bands of the bare SSH model by inspecting their parity under inversion ($\mathcal{I}$). At $k=\{\Gamma, \mathrm{X}\}$, the noninteracting Hamiltonian is
\al{
    \mathrm{H}(\Gamma) &= \pmat{0 & v+w \\ v+w & 0} &%
    \mathrm{H}(\mathrm{X}) &= \pmat{0 & v-w \\ v-w & 0}\,,
}
when taken in the $\acc{\ket{k,A,\sigma},\ket{k,B,\sigma}}$ basis. The eigenproblem is trivial:
\begin{equation}
	\begin{aligned}
        \mathrm{H}(\Gamma)\frac{1}{\sqrt{2}}\p{\ket{\Gamma,A,\sigma} \pm \ket{\Gamma,B,\sigma}} &= \cro{(v+w)\pm(v+w)}\frac{1}{\sqrt{2}}\p{\ket{\Gamma,A,\sigma} \pm \ket{\Gamma,B,\sigma}} \\
        \mathrm{H}(\mathrm{X})\frac{1}{\sqrt{2}}\p{\ket{\mathrm{X},A,\sigma} \pm \ket{\mathrm{X},B,\sigma}} &= \cro{(v-w)\pm(v-w)}\frac{1}{\sqrt{2}}\p{\ket{\mathrm{X},A,\sigma} \pm \ket{\mathrm{X},B,\sigma}}\,.
	\end{aligned}
\end{equation}
Additionally, both of these high-symmetry points are inversion symmetric, so
\al{
    \mathcal{I}\frac{1}{\sqrt{2}}\p{\ket{\Gamma/\mathrm{X},A,\sigma} \pm \ket{\Gamma/\mathrm{X},B,\sigma}} = \pm\frac{1}{\sqrt{2}}\p{\ket{\Gamma/\mathrm{X},A,\sigma} \pm \ket{\Gamma/\mathrm{X},B,\sigma}}\,.
}
Given $v,w>0$, the irreps of the noninteracting bands are assigned as in Fig. \ref{fig:SSH_bands}.
\begin{figure}[H]
    \centering
    \resizebox{0.4\linewidth}{!}{\begin{tikzpicture}
    \pgfplotsset{
        compat=1.17,
        set layers,
    }
    \pgfplotsset{
        spectral axis/.style={
            width=0.6*9cm,
            height=0.6*7cm,
            scale only axis,
        },
    }

    \def\v{1}
    \def\w{0.5}

    \pgfmathsetmacro{\DeltaLeft}{\v + \w}
    \pgfmathsetmacro{\DeltaRight}{abs(\v - \w)}


    \begin{axis}[
        spectral axis,
        name=first plot,
        xmin=0,
        xmax=pi,
        xtick={0,pi},
        ytick=\empty,
        xticklabels={\LARGE$\Gamma$, \LARGE$\mathrm{X}$},
        xlabel={{$k$}},
        ylabel={{$E$}},
        ylabel style={rotate=-90},
        label style={font=\LARGE},
        axis line style={line width=1.5pt},
        clip=false,
    ]

        \addplot[
            domain=0:pi,
            samples=100,
            color=COLOR1,
            line width=2pt,
            axis equal,
        ]
        {sqrt(\v^2 + \w^2 + 2*\v*\w*cos(deg(x)))};

        \addplot[
            domain=0:pi,
            samples=100,
            color=COLOR2,
            line width=2pt,
        ]
        {-sqrt(\v^2 + \w^2 + 2*\v*\w*cos(deg(x)))};

        \draw[COLOR1, fill]
            (0,\DeltaLeft) circle (2pt)
            node[left] {\Large$\Gamma_1^+$};

        \draw[COLOR2, fill]
            (0,-\DeltaLeft) circle (2pt)
            node[left] {\Large$\Gamma_3^-$};

        \draw[COLOR1, fill]
            (pi,\DeltaRight) circle (2pt)
            node[right] {\Large$\mathrm{X}_1^+[\mathrm{X}_3^-]$};

        \draw[COLOR2, fill]
            (pi,-\DeltaRight) circle (2pt)
            node[right] {\Large$\mathrm{X}_3^-[\mathrm{X}_1^+]$};

    \end{axis}

\end{tikzpicture}}
    \caption{Bands of the SSH chain with irreps identified at high-symmetry points. Differing values when $w>v$ are indicated in brackets.}
    \label{fig:SSH_bands}
\end{figure}

\section{Insulating Ground States} \label{sec:GS} 

The ground states of the SSH+HK model can be derived using simple arguments \cite{zhao_failure_2023, mohamadi_emergence_2025}. First, two distinct insulating phases are identified as a function of interaction strength ($U$) at half-filling ($\nu=1/2$). Then, it is shown that the quarter-filled ($\nu=1/4$) system can become a Mott insulator when interactions are sufficiently large.

Since the Hamiltonian is $k$-local, it is split into blocks
\al{
    \op{H}_k &= \epsilon_{k,+}\op{n}_{k,+} + \epsilon_{k,-}\op{n}_{k,-} + U\op{n}_{k,\uparrow}\op{n}_{k,\downarrow}\,,
}
where missing indices are implicitly summed over. Hence, at fixed filling, the chemical potential and the ground state of the whole chain are easily obtained by solving this simple interacting Hamiltonian. The energetic scales of the noninteracting Hamiltonian are set by the gap ($\Delta$), the bandwidth ($\mathrm{BW}$) and the total bandwidth ($\mathrm{TW}$) as defined in Fig. \ref{fig:BandMarkers}.

\begin{figure}[H]
\centering
\resizebox{0.33\linewidth}{!}{


\begin{tikzpicture}\pgfplotsset{compat=1.18}

\def\v {1};
\def\w {0.3};

\pgfmathsetmacro{\DeltaLeft}{\v + \w};
\pgfmathsetmacro{\DeltaRight}{abs(\v - \w)};

\begin{axis}[
    xtick=\empty,
    ytick=\empty,
    xmin=-pi, xmax=pi,
    ymin=-\v-\w-0.3, ymax=\v+\w+0.3,
    clip=false
]

\addplot [
    domain=-pi:pi,
    samples=100,
    color=COLOR1,
    line width = 2pt,
]
{sqrt( \v^2 + \w^2 + 2*\v*\w*cos(deg(x)) )};

\addplot [
    domain=-pi:pi,
    samples=100,
    color=COLOR2,
    line width = 2pt,
]
{-sqrt( \v^2 + \w^2 + 2*\v*\w*cos(deg(x)) )};




\draw[dashed] (-pi,\DeltaLeft) -- (pi,\DeltaLeft);
\draw[dashed] (-pi,\DeltaRight) -- (pi,\DeltaRight);
\draw[dashed] (-pi,-\DeltaRight) -- (pi,-\DeltaRight);
\draw[dashed] (-pi,-\DeltaLeft) -- (pi,-\DeltaLeft);


\draw[thick, <->] (-pi+0.75,\DeltaRight) -- (-pi+0.75,-\DeltaRight) node[midway, left] {\Large$\Delta$};
\draw[thick, <->] (-pi+2.25,\DeltaLeft) -- (-pi+2.25,-\DeltaLeft) node[midway, left] {\Large$\mathrm{TW}$};
\draw[thick, <->] (-pi+4,-\DeltaRight) -- (-pi+4,-\DeltaLeft) node[midway, left] {\Large$\mathrm{BW}$};


\node at (0,-\DeltaLeft) {\Large$\boldsymbol{\uparrow\downarrow}$};

\node at (-pi+0.3,-\DeltaRight-0.025) {\Large$\boldsymbol{\uparrow\downarrow}$};
\node at (pi-0.3,-\DeltaRight-0.025) {\Large$\boldsymbol{\uparrow\downarrow}$};

\node at (-pi/2+0.15,-\DeltaRight/2-\DeltaLeft/2-0.1) {\Large$\boldsymbol{\uparrow\downarrow}$};
\node at (+pi/2-0.15,-\DeltaRight/2-\DeltaLeft/2-0.1) {\Large$\boldsymbol{\uparrow\downarrow}$};

\end{axis}

\end{tikzpicture}

}
\caption{Energy scales of the noninteracting SSH chain. The bands of bandwidth $\mathrm{BW}$ are separated by the gap $\Delta$ thus constituting a two-band model of total width $\mathrm{TW}$.}
\label{fig:BandMarkers}
\end{figure}

\subsection{Half-Filling}\label{subsec:GSHalfFilling} 

In the absence of interactions, the model is a band insulator at $\nu=1/2$ with the typical Fermi sea ground state \cite{OG_SSH, OG_SSH_2}. In this state, $n_{k,\sigma} = 1$ (see Fig. 1b of Letter). As $U$ is increased, the interaction $U\op{n}_{k,-,\uparrow}\op{n}_{k,-,\downarrow}$ contributes to the energy of states. However, as long as the interaction remains smaller than the gap $U<\Delta$, it is more energetically favorable for electrons to remain in the lower band. The ground state of the interacting band insulator (BI+U) defined by $\nu=1/2$ and $U<\Delta$ is
\al{
    \ket{\Omega_{\text{BI+U}}} = \prod_{k}\op{c}^\dagger_{k-\uparrow}\op{c}^\dagger_{k-\downarrow}\ket{0}\,.  \label{eq:GS_case1}
}

As the interaction is further increased, the $k$-points for which the interaction exceeds the $k$-local gap possess a ground state of the form $\op{c}^\dagger_{k,+,\sigma}\op{c}^\dagger_{k,-,\sigma}\ket{0}$ (see Fig. \ref{fig:transitionBS}), where both electrons have the same spin because the interaction term does not contain $\op{n}_{k,-,\sigma}\op{n}_{k,+,\sigma}$. This state is metallic in the same way that a partially filled band is.

\begin{figure}[H]
\centering
\resizebox{0.33\linewidth}{!}{


\begin{tikzpicture}\pgfplotsset{compat=1.18}

\def\v {1};
\def\w {0.3};

\def \wiggleamplitude {0.3pt}; 
\def \wiggleperiod {3pt}; 

\pgfmathsetmacro{\DeltaLeft}{\v + \w};
\pgfmathsetmacro{\DeltaRight}{abs(\v - \w)};

\begin{axis}[
    xtick=\empty,
    ytick=\empty,
    xmin=-pi, xmax=pi,
    ymin=-\v-\w-0.3, ymax=\v+\w+0.3,
    clip=false
]

\addplot [
    domain=-pi:pi,
    samples=100,
    color=COLOR1,
    line width = 2pt,
]
{sqrt( \v^2 + \w^2 + 2*\v*\w*cos(deg(x)) )};

\addplot [
    domain=-pi:pi,
    samples=100,
    color=COLOR2,
    line width = 2pt,
]
{-sqrt( \v^2 + \w^2 + 2*\v*\w*cos(deg(x)) )};







\draw[COLOR3, very thick] (-pi,-\DeltaRight) -- (-pi+1.3,-\DeltaRight);
\draw[COLOR3, very thick] (-pi,\DeltaRight+0.18) -- (-pi+1.3,\DeltaRight+0.18);
\draw[very thick, <->, COLOR3, decorate, decoration={snake, amplitude=\wiggleamplitude, segment length=\wiggleperiod},] (-pi+1.1,-\DeltaRight) -- (-pi+1.1,\DeltaRight+0.18) node [midway, right] {\Large $U$};


\node at (0,-\DeltaLeft) {\Large\color{black} $\boldsymbol{\uparrow}\boldsymbol{\downarrow}$};

\node at (-pi+0.3,-\DeltaRight-0.025) {\Large\color{black} $\boldsymbol{\uparrow}$};
\node at (pi-0.3,-\DeltaRight-0.025) {\Large\color{black} $\boldsymbol{\downarrow}$};

\node at (-pi/2+0.15,-\DeltaRight/2-\DeltaLeft/2-0.1) {\Large\color{black} $\boldsymbol{\uparrow}\boldsymbol{\downarrow}$};
\node at (+pi/2-0.15,-\DeltaRight/2-\DeltaLeft/2-0.1) {\Large\color{black} $\boldsymbol{\uparrow}\boldsymbol{\downarrow}$};


\node at (-pi+0.3,+\DeltaRight+0.025) {\Large\color{black} $\boldsymbol{\uparrow}$};
\node at (pi-0.3,+\DeltaRight+0.025) {\Large\color{black} $\boldsymbol{\downarrow}$};

\draw[very thick, decorate, decoration={snake, amplitude=\wiggleamplitude, segment length=\wiggleperiod}, COLOR3] 
    (0,-\DeltaLeft-0.02) ellipse (0.35 and 0.22);
\draw[very thick, decorate, decoration={snake, amplitude=\wiggleamplitude, segment length=\wiggleperiod}, COLOR3] 
    (-pi/2+0.15,-\DeltaRight/2-\DeltaLeft/2-0.1-0.02) ellipse (0.35 and 0.22);
\draw[very thick, decorate, decoration={snake, amplitude=\wiggleamplitude, segment length=\wiggleperiod}, COLOR3] 
    (+pi/2-0.15,-\DeltaRight/2-\DeltaLeft/2-0.1-0.02) ellipse (0.35 and 0.22);

\end{axis}

\end{tikzpicture}

}
\caption{Graphical representation of the transitional metallic phase between BI+U and HFMI ($\Delta\leq U\leq\mathrm{TW}$). Regions where the $k$-local gap in the noninteracting band structure is less than $U$ have a local ground state with an electron in the upper and the lower band. Conversely, the ground state remains identical to the noninteracting case when the $k$-local gap is larger.}
\label{fig:transitionBS}
\end{figure}

Finally, as the interaction strength is raised above the total band structure width ($\mathrm{TW}$), both bands will be entirely singly occupied with the constraint that the spin at each $k$-point be of one given variety (see Fig. 1b in Letter). Hence, the degenerate ground states at each $k$ are
\al{
    \mathscr{H}_0^{(\text{HFMI})}(k) = \acc{\op{c}^\dagger_{k,+,\uparrow}\op{c}^\dagger_{k,-,\uparrow}\ket{0}, \op{c}^\dagger_{k,+,\downarrow}\op{c}^\dagger_{k,-,\downarrow}\ket{0}}\,.
}
Since either ground state contains a copy of a full SSH chain, they are equivalent to a trivial atomic limit. To prove this statement, note that
\al{
    \pmat{\op{c}^\dagger_{k,+,\sigma}\\\op{c}^\dagger_{k,-,\sigma}} = \mathrm{U}(k)\pmat{\op{c}^\dagger_{k,A,\sigma}\\\op{c}^\dagger_{k,B,\sigma}}\,
}
where
\al{
    \mathrm{U}(k) = \frac{1}{\sqrt{2}}\pmat{
    \Exp{i\phi(k)} & 1\\
    \Exp{i\phi(k)} & -1\\
    }
}
reveals that
\al{
    \op{c}^\dagger_{k,+,\sigma}\op{c}^\dagger_{k,-,\sigma}\ket{0} = \operatorname{det}(\mathrm{U})\op{c}^\dagger_{k,A,\sigma}\op{c}^\dagger_{k,B,\sigma}\ket{0} \sim \op{c}^\dagger_{k,A,\sigma}\op{c}^\dagger_{k,B,\sigma}\ket{0}\,.
}

\subsection{Quarter-Filling} 

Conversely, at $\nu = 1/4$, the model is metallic at $U=0$ as the lower band is half-filled. As $U$ is increased, electrons spread out across the lower band such that the total energy is minimized. However, as $U$ reaches the bandwidth ($\mathrm{BW}$), the lowest-energy configurations are those in which electrons are distributed such that the lower band is entirely singly occupied (Fig. 1b of the Letter). The resulting phase can be characterized as a Mott insulator as the metallic nature of the noninteracting system gives way entirely to a correlated insulator. The degenerate ground states at each $k$ are simply
\al{
    \mathscr{H}_0^{(\text{QFMI})}(k) = \acc{\op{c}^\dagger_{k,-,\uparrow}\ket{0}, \op{c}^\dagger_{k,-,\downarrow}\ket{0}
    }\,.\label{eq:GS_case3}
}

\section{Interacting Single-Particle Picture} \label{sec:ISSP} 

\subsection{Definitions and conventions}\label{subsec:ISSP_GF}

\subsubsection{Zero-temperature Green's function}

Given the exact many-body ground states of the insulating phases, the Green's function can be computed directly. At zero temperature, we may use the following Lehmann decomposition \cite{PyQCM}:
\al{
    G_{n,\sigma}(k,z) &= \frac{1}{g}\sum_{d=1}^{g}\bra{\Omega_d(k)}\cro{
        \op{c}_{k,n,\sigma}\frac{\mathbbm{1}}{z - \op{H}_k + E_\Omega}
    \op{c}_{k,n,\sigma}^\dagger + \op{c}_{k,n,\sigma}^\dagger\frac{\mathbbm{1}}{z + \op{H}_k - E_\Omega}
    \op{c}_{k,n,\sigma}}\ket{\Omega_d(k)}\,,
}
where $k$-locality has been used. In analytical calculations, it is often practical to split the Green's function into its \enquote{electron} and \enquote{hole} parts:
\begin{align}
    G_{n,\sigma}^{(e)}(k,z) &= \frac{1}{g}\sum_{d=1}^{g}\frac{\bra{\Omega_d(k)}
        \op{c}_{k,n,\sigma}\op{c}_{k,n,\sigma}^\dagger\ket{\Omega_d(k)}}{z - E^{(e)}_{k,n,\sigma;d} + E_\Omega}\\
    G_{n,\sigma}^{(h)}(k,z) &= \frac{1}{g}\sum_{d=1}^{g}\frac{\bra{\Omega_d(k)}
        \op{c}_{k,n,\sigma}^\dagger\op{c}_{k,n,\sigma}\ket{\Omega_d(k)}}{z + E^{(h)}_{k,n,\sigma;d} - E_\Omega}\,.
\end{align}
The energies of the single-particle excitations are
\al{
    E^{(e)}_{k,n,\sigma;d}\equiv \bra{\Omega_{k;d}}\op{c}_{k,n,\sigma}\op{H}_k\op{c}_{k,n,\sigma}^\dagger\ket{\Omega_{k;d}}\\
    E^{(h)}_{k,n,\sigma;d}\equiv \bra{\Omega_{k;d}}\op{c}_{k,n,\sigma}^\dagger\op{H}_k\op{c}_{k,n,\sigma}\ket{\Omega_{k;d}}
}
and the weight of each associated pole is defined by
\begin{align}
    p^{(e)}_{k,n,\sigma;d}&\equiv\frac{1}{g}\bra{\Omega_d(k)}
    \op{c}_{k,n,\sigma}\op{c}_{k,n,\sigma}^\dagger\ket{\Omega_d(k)}\\
    p^{(h)}_{k,n,\sigma;d}&\equiv\frac{1}{g}\bra{\Omega_d(k)}
    \op{c}_{k,n,\sigma}^\dagger\op{c}_{k,n,\sigma}\ket{\Omega_d(k)}\\
    1 &= \sum_{d=1}^{g}\p{p^{(e)}_{k,n,\sigma;d} + p^{(h)}_{k,n,\sigma;d}}\,,
\end{align}
where the last equality holds due to canonical anticommutation. These weights represent the statistical contribution of each ground state to the dynamics. All of these quantities can be calculated individually and are later recombined into the full Green's function:
\al{
    G_{n,\sigma}(k,z) = \sum_{d}\cro{ \frac{p_{k,n,\sigma;d}^{(e)}}{z - E^{(e)}_{k,n,\sigma;d} + E_\Omega} + \frac{p_{k,n,\sigma;d}^{(h)}}{z + E^{(h)}_{k,n,\sigma;d} - E_\Omega}}\,,
}
which also offers a particularly clear insight into the structure of the Green's function.

\subsubsection{Single-particle reduced density matrix}

Although the 1RDM is defined as a ground state expectation value, all necessary information is encoded within the Green's function. At zero temperature, it is well known that
\al{
    \gamma_{\mu\nu} = \int_{\mathcal{C}_<}\frac{\D{z}}{2\pi i}G_{\mu\nu}(z)\,,
}
where $\mathcal{C}_{<}$ is a counterclockwise contour enclosing all poles under the chemical potential. Since the Green's function is a series of simple poles \cite{rickayzen1980green}, the 1RDM in the band basis is the sum of weights for all the poles under the chemical potential. In this case, this corresponds to summing all weights associated with hole excitations
\al{
    \op{\gamma}_{\sigma}(k) = \sum_{d}\p{p^{(h)}_{k,+,\sigma;d}\ketbra{k,+,\sigma}{k,+,\sigma} + p^{(h)}_{k,-,\sigma;d}\ketbra{k,-,\sigma}{k,-,\sigma}}\,.
}

\subsection{Calculations}

\subsubsection{BI+U}

In BI+U, the ground state at each $k$ is $\op{c}^\dagger_{k,-,\uparrow}\op{c}^\dagger_{k,-,\downarrow}\ket{0}$ with an energy of
\al{
    E_\Omega = 2\epsilon_{k,-} + U\,.
}
The electronic part only has a contribution for $n=+$ since the lower band is fully occupied, so
\al{
    p_{n}^{(e)} = \delta_{n+}
}
with an excitation energy of
\al{
    E_n^{(e)} = 2\epsilon_{k,-} + \epsilon_{k,+} + 2U\,.
}
Similarly, the hole part only contributes when $n=-$, such that
\al{
    p_{n}^{(h)} = \delta_{n-}
}
and
\al{
    E_n^{(h)} = \epsilon_{k,-}\,.
}
Hence, we obtain the Green's function for the BI+U phase:
\al{
    G_{n,\sigma}(k,z) = \frac{\delta_{n,+}}{z - (\epsilon_{k,+} + U)} + \frac{\delta_{n,-}}{z - (\epsilon_{k,-} + U)}\,.
}
The corresponding 1RDM is then
\al{
    \op{\gamma}_{\sigma}(k) = \ketbra{k,-,\sigma}{k,-,\sigma}\,.
}

\subsubsection{QFMI}

For QFMI, the ground states to consider at each $k$-point are $\{ \op{c}^\dagger_{k,-,\uparrow}\ket{0}, \op{c}^\dagger_{k,-,\downarrow}\ket{0} \}$ such that the relevant weights are:
\al{
    p_n^{(e)} &= \delta_{n,+} + \frac{\delta_{n,-}}{2}\\
    p_n^{(h)} &= \frac{\delta_{n,-}}{2}
}
The ground state energy is simply
\al{
    E_\Omega = \epsilon_{-}
}
while the excited energies for added and removed particles are
\al{
    E_{n,\sigma;d}^{(e)} &= E_\Omega + \epsilon_n + \delta_{\sigma, -\sigma_{d}}U\\
    E_{-,\sigma;d}^{(h)} &= 0\,,
}
where $\sigma_d$ is the spin associated with the degenerate ground state labeled by $d$. The Green's function reads
\al{
    G_{n,\sigma}(k,z) = &\frac{\delta_{n,+}}{2}\cro{\frac{1}{z - (\epsilon_{k,+} + U)} + \frac{1}{z - \epsilon_{k,+}}}\notag\\
    + &\frac{\delta_{n,-}}{2}\cro{\frac{1}{z - (\epsilon_{k,-} + U)} + \frac{1}{z - \epsilon_{k,-}}}\,,
}
with its associated 1RDM
\al{
    \op{\gamma}_{\sigma}(k) = \frac{1}{2}\ketbra{k,-,\sigma}{k,-,\sigma}\,.
}

\subsubsection{HFMI}

The two ground states of the HFMI are $\{ \ket{+_\uparrow-_\uparrow}, \ket{+_\downarrow-_\downarrow} \}$ at each $k$. Focusing first on the weights,
\al{
    p_n^{(e)} = \frac{1}{2} = p_n^{(h)}
}
and the energy of the ground states
\al{
    E_\Omega = \epsilon_+ + \epsilon_-\,,
}
we may now calculate the associated excitation energies:
\al{
    E_{n}^{(e)} &= E_\Omega + \epsilon_n + 2U\\
    E_n^{(h)} &= E_\Omega - \epsilon_n\,.
}
The full Green's function for HFMI is then
\al{
    G_{n,\sigma}(k,z) = &\frac{\delta_{n,+}}{2}\cro{\frac{1}{z - (\epsilon_{k,+} + 2U)} + \frac{1}{z - \epsilon_{k,+}}}\notag\\
    + &\frac{\delta_{n,-}}{2}\cro{\frac{1}{z - (\epsilon_{k,-} + 2U)} + \frac{1}{z - \epsilon_{k,-}}}\,,
}
with its associated 1RDM
\al{
    \op{\gamma}_{\sigma}(k) = \frac{1}{2}\ketbra{k,-,\sigma}{k,-,\sigma} + \frac{1}{2}\ketbra{k,+,\sigma}{k,+,\sigma}\,.
}

\section{Effective Atomic Limits} \label{sec:EAL} 

Symmetry-resolved spectral weight can be used to extend elementary band representations (EBRs) to interacting electrons. In this section, possible atomic limits are constructed from the occupied spectral weight in all phases.

From the irreps at high-symmetry points, all distinct EBRs can be determined with the BANDREP utility of the Bilbao Crystallographic Server \cite{Bradlyn2017TQC, GraphTheoryMaia}. In this case, the relevant EBRs and atomic limits are listed in Tab. \ref{tab:EBRs}.

\begin{table}[H]
\centering
\setlength{\tabcolsep}{25pt}
\renewcommand{\arraystretch}{1.2}
\begin{tabularx}{0.5\columnwidth}{l l}
\toprule
EBR & Atomic Limit \\ \midrule
$(\Gamma_1^{+},\mathrm{X}_1^{+})$ & $(\mathrm{A}_g \uparrow G(1))_{1a}$ \\
$(\Gamma_1^{+},\mathrm{X}_3^{-})$ & $(\mathrm{A}_g \uparrow G(1))_{1b}$ \\
$(\Gamma_3^{-},\mathrm{X}_1^{+})$ & $(\mathrm{B}_{3u} \uparrow G(1))_{1b}$ \\
$(\Gamma_3^{-},\mathrm{X}_3^{-})$ & $(\mathrm{B}_{3u} \uparrow G(1))_{1a}$ \\\bottomrule
\end{tabularx}
\caption{Relevant EBRs and atomic limits of space group Pmmm.}
\label{tab:EBRs}
\end{table}

\subsection{BI+U and QFMI}

The effective atomic limits of BI+U and QFMI differ only by an overall weight of two in their occupied spectral weight, so we may analyse these simultaneously. Whereas the occupied spectral weight at $w<v$ is a multiple of $(\Gamma_3^{-},\mathrm{X}_3^{-})$, the decomposition at $w>v$ is rather $(\Gamma_3^{-},\mathrm{X}_1^{+})$. Both of these form multiples of a whole EBR and naturally correspond to different atomic limits (Tab. \ref{tab:EBRs}).

The half-weighted nature of the QFMI is naturally interpreted from the half-weight of the spectral weight due to quarter-filling. In general, we adopt the notion that spectral weight of an interacting system must be decomposed into combinations of EBRs with an overall weighting factor.

\subsection{HFMI}

In the case of HFMI, the occupied spectral weight, independently of $w$, is written as $\frac{1}{2}(\Gamma_1^{+}\oplus\Gamma_3^{-},\mathrm{X}_1^{+}\oplus\mathrm{X}_3^{-})$. This is the irrep decomposition of the fully filled SSH chain with a factor of one-half due to the total number of electrons corresponding to half-filling.

There are two possible pairings of irreps into EBRs that completely decompose the occupied spectral weight. If one starts with $(\Gamma_1^{+},\mathrm{X}_1^{+})$, the remaining irreps are paired as $(\Gamma_3^{-},\mathrm{X}_3^{-})$ which gives the atomic limit $(\mathrm{A}_g \uparrow G(1))_{1a}\oplus(\mathrm{B}_{3u} \uparrow G(1))_{1a}$. Conversely, we could have started with $(\Gamma_1^{+},\mathrm{X}_3^{-})$ which requires $(\Gamma_3^{-},\mathrm{X}_1^{+})$, thus yielding $(\mathrm{A}_g \uparrow G(1))_{1b}\oplus(\mathrm{B}_{3u} \uparrow G(1))_{1b}$ as an atomic limit.

\section{Spin-Resolved Bulk Polarization} \label{sec:Pol} 

In this section, the spin-resolved bulk polarization is derived in the spirit of Resta \cite{RestaCanonicalPol}, yielding a form similar to related work \cite{mohamadi_emergence_2025} and is then calculated for all phases of the spin-symmetric SSH+HK model.

\subsection{Derivation}

Given the extensive position operator in spin sector $\sigma$
\al{
    \op{X}_\sigma \equiv \sum_{j}x_{j}\op{n}_{j,\sigma}\,,
}
its correct expectation value for periodic boundary conditions (PBC) is \cite{RestaCanonicalPol}
\al{
    \avg{\op{X}_\sigma}_{\text{PBC}} \equiv  \frac{L}{2\pi}\Im{\ln\avg{\Exp{i2\pi\op{X}_\sigma/L}}}\,.
}
As elegantly argued in \cite{RestaCanonicalPol}, the bulk polarization per unit length $a=1$ is obtained in the thermodynamic limit as
\al{
    P_{\sigma} = \lim_{L\to\infty}\frac{e\avg{\op{X}_\sigma}_{\text{PBC}}}{L}\,,
}
which is defined modulo $e$. We may now make use of the $k$-factored structure of the ground state manifold in HK models to relate this formula to a convenient winding number formulation. The challenging expression to evaluate is the thermodynamic limit of the expectation value. The following identities of the twist operator are of crucial importance \cite{BradlynIraolaNotes}:
\al{
    \Exp{i2\pi\op{X}_\sigma/L}\op{c}^{\dagger}_{k,a,\sigma}\Exp{-i2\pi\op{X}_\sigma/L} &= \op{c}^{\dagger}_{k+\delta{k},a,\sigma} & \Exp{i2\pi\op{X}_\sigma/L}\op{c}^{\dagger}_{k,a,-\sigma}\Exp{-i2\pi\op{X}_\sigma/L} &= \op{c}^{\dagger}_{k,a,-\sigma}\,,\label{eq:TwistAndShout}
}
where $\delta{k} = 2\pi/L$.

\subsubsection{Non-Degenerate Ground State}

In the case of a non-degenerate ground state $\ket{\Omega} = \bigotimes_{k}\ket{\Psi(k)}$, calculating this polarization is straightforward. We start by noting that the expectation value reduces to
\begin{equation}\begin{aligned}
    \bra{\Omega}\Exp{i2\pi\op{X}_\sigma/L}\ket{\Omega} &= \bra{\Omega_\sigma}\Exp{i2\pi\op{X}_\sigma/L}\ket{\Omega_\sigma}\\
    &= \bigotimes_{k'}\bra{\Psi_\sigma(k')}\bigotimes_{k}\ket{\Psi_\sigma(k + \delta{k})}\\
    &= \bigotimes_{k'}\bra{\Psi_\sigma(k')}\bigotimes_{k}\cro{\ket{\Psi_\sigma(k)} + \del_{k}\ket{\Psi_\sigma(k)}\delta{k} + \mathcal{O}(\delta{k}^2)}\\
    &= \prod_k \cro{1 + \bra{\Psi_\sigma(k)}\del_{k}\ket{\Psi_\sigma(k)}\delta{k} + \mathcal{O}(\delta{k}^2)}\,,
\end{aligned}\end{equation}
where we have used the $k$-locality of the ground state $\ket{\Psi(k)} = \ket{\Psi_{\uparrow}(k)}\otimes\ket{\Psi_{\downarrow}(k)}$. Taking the thermodynamic limit of the logarithm gives a clean expression:
\begin{equation}\begin{aligned}
    \lim_{L\to\infty}\ln\avg{\Exp{i2\pi\op{X}_\sigma/L}} &= \lim_{L\to\infty}\sum_k \ln\cro{1 + \bra{\Psi_\sigma(k)}\del_{k}\ket{\Psi_\sigma(k)}\delta{k} + \mathcal{O}(\delta{k}^2)}\\
    &= \lim_{L\to\infty}\sum_k \cro{\bra{\Psi_\sigma(k)}\del_{k}\ket{\Psi_\sigma(k)}\delta{k} + \mathcal{O}(\delta{k}^2)}\\
    &= \int_{\text{1BZ}} \bra{\Psi_\sigma(k)}\del_{k}\ket{\Psi_\sigma(k)}\D{k}\,.
\end{aligned}\end{equation}
Because this term is purely imaginary, the spin-resolved polarization evaluates exactly to:
\al{
    P_\sigma = \frac{-ie}{2\pi}\int_{\text{1BZ}} \bra{\Psi_\sigma(k)}\del_{k}\ket{\Psi_\sigma(k)}\D{k} ~(\operatorname{mod}e)\,.\label{eq:PolNonDeg}
}

\subsubsection{Degenerate Ground States}

In the case of QFMI and HFMI, the ground state degeneracy at each $k$-point is twofold and is faithfully labelled by spin. Therefore, their equilibrium density operators are of the following form:
\al{
    \rho = \bigotimes_{k}\frac{1}{2}\cro{\ketbra{\Psi_\uparrow(k)}{\Psi_\uparrow(k)} + \ketbra{\Psi_\downarrow(k)}{\Psi_\downarrow(k)}}\,.
}
Taking the $\sigma=\uparrow$ sector for example, the expectation value reads as:
\begin{equation}\begin{aligned}
    \Tr{\Exp{i2\pi\op{X}_\uparrow/L}\rho} &= \Tr{\Exp{i2\pi\op{X}_\uparrow/L}\bigotimes_{k}\frac{1}{2}\cro{\ketbra{\Psi_\uparrow(k)}{\Psi_\uparrow(k)} + \ketbra{\Psi_\downarrow(k)}{\Psi_\downarrow(k)}}}\\
    &= \Tr{\bigotimes_{k}\frac{1}{2}\cro{\ketbra{\Psi_\uparrow(k+\delta{k})}{\Psi_\uparrow(k)} + \ketbra{\Psi_\downarrow(k)}{\Psi_\downarrow(k)}}}\\
    &= \Tr{\bigotimes_{k}\frac{1}{2}\cro{\ketbra{\Psi_\uparrow(k)}{\Psi_\uparrow(k)} + \ketbra{\Psi_\downarrow(k)}{\Psi_\downarrow(k)} + \ketbra{\del_{k}\Psi_\uparrow(k)}{\Psi_\uparrow(k)}\delta k + \mathcal{O}(\delta k^2)}}\,.
\end{aligned}\end{equation}
In the last equality, we have used the Taylor expansion of $\ket{\Psi_\uparrow(k)}$ in powers of $\delta k$. The use of a series expansion is well justified since taking the thermodynamic limit sends $\delta{k}\to0$ as $\frac{1}{L}$ and promotes the wavevector to a continuous variable allowing for a well-defined derivative. In particular, this manipulation becomes convenient when evaluating the thermodynamic limit of the logarithm, as the limit also allows the density operator to be split into independent $k$-sectors.
\begin{equation}\begin{aligned}
    \lim_{L\to\infty}\ln\avg{\Exp{i2\pi\op{X}_\sigma/L}} &= \lim_{L\to\infty}\sum_k \ln\p{1 + \frac{1}{2}\bra{\Psi_\uparrow(k)}\del_{k}\ket{\Psi_\uparrow(k+\delta{k})}\delta{k} + \mathcal{O}(\delta{k}^2)}\\
    &= \lim_{L\to\infty}\sum_k \cro{\frac{1}{2}\bra{\Psi_\sigma(k)}\del_{k}\ket{\Psi_\sigma(k)}\delta{k} + \mathcal{O}(\delta{k}^2)}\\
    &= \frac{1}{2}\int_{\text{1BZ}} \bra{\Psi_\sigma(k)}\del_{k}\ket{\Psi_\sigma(k)}\D{k}\,.
\end{aligned}\end{equation}
Here, the Riemann sum is converted to an integral after using the series expansion of the logarithm $\ln x = x + \mathcal{O}(x^2)$. Finally, the spin-resolved polarization is
\al{
    P_\sigma = \frac{-ie}{4\pi}\int_{\text{1BZ}} \bra{\Psi_\sigma(k)}\del_{k}\ket{\Psi_\sigma(k)}\D{k} ~(\operatorname{mod}e)\,,\label{eq:PolDeg}
}
where degeneracy effectively halves the contribution from the winding.

\subsection{Calculations}

To simplify the following calculations, we recall that
\al{
    \op{c}^\dagger_{k,\pm,\sigma} = \frac{1}{\sqrt{2}}\p{\Exp{i\phi(k)}\op{c}^\dagger_{k,A,\sigma} \pm \op{c}^\dagger_{k,B,\sigma}}\,
}
such that
\al{
    i\del_k\op{c}^\dagger_{k,\pm,\sigma} = \frac{-1}{\sqrt{2}}\del_k\phi(k)\Exp{i\phi(k)}\op{c}^\dagger_{k,A,\sigma}\,.
}
Since
\al{
    \op{c}^\dagger_{k,A,\sigma} = \frac{1}{\sqrt{2}}\Exp{-i\phi(k)}(\op{c}^\dagger_{k,+,\sigma} + \op{c}^\dagger_{k,-,\sigma})\,,
}
we may write down the following convenient form:
\al{
    \del_k\op{c}^\dagger_{k,\pm,\sigma} = \frac{i}{2}\del_k\phi(k)(\op{c}^\dagger_{k,+,\sigma} + \op{c}^\dagger_{k,-,\sigma})\,.
}
Finally, we remind the reader that depending on the parameter regime
\al{
    \int_{\text{1BZ}}\text{d}k\,\del_k\phi(k) = \sys{
        0~~,~~w<v\\
        2\pi~~,~~w>v\\
    }\,.
}

\subsubsection{BI+U}

In the case of BI+U, the ground state in the $(k,\sigma)$ sector is $\ket{\Omega_\sigma(k)} = \op{c}^\dagger_{k,-,\sigma}\ket{0}$. Hence, the matrix element follows
\begin{equation}
\begin{aligned}
	\bra{\Omega_\sigma(k)}\partial_k\ket{\Omega_\sigma(k)} &= \bra{0}\frac{i}{2}\del_k\phi(k)\op{c}_{k,-,\sigma}(\op{c}^\dagger_{k,+,\sigma} + \op{c}^\dagger_{k,-,\sigma})\ket{0} = \frac{i}{2}\del_k\phi(k)\,.
\end{aligned}
\end{equation}
Therefore, the spin-resolved winding number \eqref{eq:PolNonDeg} is
\al{
    P_{\sigma} = \frac{-ie}{4\pi}\int_{\text{1BZ}} \frac{i}{2}\del_k\phi(k)\D{k} ~(\operatorname{mod}e) =\sys{
    0~~,~~w<v\\
    e/2~~,~~w>v
    }\,.\label{eq:BIpUPol}
}
As expected, this is identical to the polarization of the spinless noninteracting SSH chain, since the spinful BI+U state contains one copy of the spinless model's ground state per spin sector.

\subsubsection{QFMI}

The ground state of QFMI is decomposed into $\ket{\Omega_\sigma(k)} = \op{c}^\dagger_{k,-,\sigma}\ket{0}$. Thus, the relevant matrix elements are
\begin{equation}
\begin{aligned}
	\bra{\Omega_\sigma(k)}i\partial_k\ket{\Omega_\sigma(k)} &= \frac{-1}{2}\del_k\phi(k)\,.
\end{aligned}
\end{equation}
Using the formula derived above \eqref{eq:PolDeg} for a degenerate ground state, the spin-resolved polarization is calculated:
\al{
    P_{\sigma} = \frac{e}{2\pi}\frac{1}{2}\cdot1\cdot\int_{\text{1BZ}}\D{k}\frac{-1}{2}\del_k\phi(k) ~(\operatorname{mod}e) =\sys{
    0~~,~~w<v\\
    e/4~~,~~w>v
    }\,.
}
Compared to BI+U \eqref{eq:BIpUPol}, the half-valued polarization here is due to QFMI containing only one copy of the spinless SSH chain's ground state split evenly across both spin projections at each $k$-point.

\subsubsection{HFMI}

The ground state of HFMI factors into $\ket{\Omega_\sigma(k)} = \op{c}^\dagger_{k,+,\sigma}\op{c}^\dagger_{k,-,\sigma}\ket{0}$. However, recalling the discussion in Sec. \ref{subsec:GSHalfFilling}, we know that $\ket{\Omega_\sigma(k)} \sim \op{c}^\dagger_{k,A,\sigma}\op{c}^\dagger_{k,B,\sigma}\ket{0}$ which is a trivial atomic limit, i.e., the orbitals of the tight binding model itself. Given the position of the orbitals within the unit cell, the spin-resolved polarization of HFMI must vanish
\al{
    P_{\sigma} = 0\,.
}

\section{1RDM Geometry} \label{sec:QV} 

First, we briefly cover covariant differentiation. Then, the notion of distance in one-body state space is constructed via the quantum metric.

\subsection{Covariant Differentiation} 

The 1RDM defines a map between a parameter manifold $\mathcal{M}$ and the space of hermitian positive semi-definite operators $\mathcal{D}^{+}$ \cite{BradlynIraolaNotes}
\al{
    \op{\gamma}: \mathcal{M} \to \mathcal{D}^{+}\,.
}
These operators are then represented in any orthonormal basis as
\al{
    \op{\gamma}(\lambda) = \gamma_{\mu\nu}(\lambda)\ketbra{\phi_\mu(\lambda)}{\phi_\nu(\lambda)}\,,
}
where $\lambda$ is a coordinate parametrization of $\mathcal{M}$. Then, the derivative of the density operator can be written in components
\al{
    \bra{\phi_\alpha}\del_\lambda\op{\gamma}\ket{\phi_\beta} = \del_\lambda\gamma_{\alpha\beta} -i\mathcal{A}_{\alpha\nu}\gamma_{\nu\beta} + i\gamma_{\alpha\nu}\mathcal{A}_{\nu\beta}\,,
}
where
\al{
    \mathcal{A}_{\mu\nu} = i\bra{\phi_\mu}\del_\lambda\ket{\phi_\nu}
}
is the non-Abelian Wilczek-Zee connection \cite{WZConnection}. Using bold symbols to denote matrix quantities, the covariant derivative
\al{
    \mathscr{D}_\lambda\boldsymbol{\gamma} = \del_\lambda\boldsymbol{\gamma} -i\cro{\boldsymbol{\mathcal{A}},\boldsymbol{\gamma}}
}
transforms as a coordinate matrix \cite{fradkin2021quantum}
\al{
    \mathscr{D}_\lambda'\boldsymbol{\gamma}' = \ve{U}\mathscr{D}_\lambda\boldsymbol{\gamma}\ve{U}^\dagger
}
provided that the connection transforms as usual:
\al{
    \boldsymbol{\mathcal{A}}' = \ve{U}\boldsymbol{\mathcal{A}}\ve{U}^\dagger - i\del_\lambda(\ve{U})\ve{U}^\dagger\,.
}
This connection can be considered as the local connection on a $U(N)$ principal bundle \cite{Jamiolkowski, Nakahara}.

\begin{figure}[H]
    \centering
    \resizebox{0.5\linewidth}{!}{\begin{tikzpicture}

\pgfmathsetmacro{\CircFac}{sqrt(3)/2}

\draw[very thick] (0,0) arc (-180:180:1);
\node at (1,0) {1BZ};

\node[circle, fill=black, inner sep=1pt] (kpoint) at (1+\CircFac,1/2) {};
\node[left] at (kpoint) {\large$k$};

\draw[very thick, dashed, COLOR3] (4,0) arc (-180:180:1); 
\draw[very thick, COLOR2] (6,0) arc (0:235:1); 
\node at (5,0) {State Space};

\node (dpoint) at (5-\CircFac,1/2) {} ;
\draw[-stealth, very thick, COLOR1] ($(dpoint) + 0.045*(1/2,\CircFac)$) -- ($(dpoint) - 0.7*(1/2,\CircFac)$) node[near end, left] {\large$\mathcal{D}_k\ve{\gamma}$};
\node[circle, fill=COLOR2, inner sep=1pt] at (dpoint) {};
\node at (dpoint) [right] {\large$\color{COLOR2}\ve{\gamma}(k)$};

\draw [-stealth, very thick] ($(kpoint) + (4pt,2pt)$) to [out=30,in=150] ($(dpoint) + (-4pt,2pt)$);

\end{tikzpicture}}
    \caption{
        Illustration of the 1RDM as a map from the 1BZ to one-body state space. For a fixed parameter regime, its image (blue) need not cover the full state space (dashed ring). The covariant derivative $\mathcal{D}_k\gamma$ (red) lies tangent to this image.
    }
    \label{fig:QuantumVolumeSchematic}
\end{figure}

\subsection{State Metric}

We now define the notion of distance on the space of 1RDMs. In this work, we use the Hilbert-Schmidt distance defined for operators as:
\al{
    \p{\op{A}, \op{B}}_{\text{HS}} = \Tr{\op{A}^\dagger\op{B}}
}
This can be used to define an infinitesimal distance element or \textit{metric} in the space of 1RDMs. If we take the derivative along a path $\lambda\in\mathcal{M}$, the line element is
\eqn{
    \text{d}{s}^2 \propto {\p{\mathscr{D}_\lambda\boldsymbol{\gamma}, \mathscr{D}_\lambda\boldsymbol{\gamma} }}\text{d}\lambda^2 = {\operatorname{Tr}\cro{\p{\mathscr{D}_\lambda\boldsymbol{\gamma}}^\dagger\mathscr{D}_\lambda\boldsymbol{\gamma}}}\text{d}\lambda^2\,,
}
where we have left the possibility for a scaling coefficient. The correct normalization is fixed by the pure state limit where the appropriate metric is the Fubini-Study metric (often referred to as the quantum geometric tensor) \cite{cheng2013quantumgeometrictensorfubinistudy, Nakahara}. To do this, let $\op{\gamma}(\lambda)=\ketbra{\psi(\lambda)}{\psi(\lambda)}$ and $\dot{\circ}\equiv\partial_\lambda\circ$. The line element is
\al{
    \text{d}s^2 = \kappa\Tr{\dot{\op{\gamma}}^2}
}
and the quantity in the trace is
\al{
    \dot{\op{\gamma}}^2 = &\ket{\dot{\psi}}\!\!\braket{\psi|\dot{\psi}}\!\!\bra{\psi} + \ket{\dot{\psi}}\!\!\braket{\psi|\psi}\!\!\bra{\dot{\psi}}\notag\\
        +&\ket{\psi}\!\!\braket{\dot{\psi}|\dot{\psi}}\!\!\bra{\psi} + \ket{\psi}\!\!\braket{\dot{\psi}|\psi}\!\!\bra{\dot{\psi}}\,.
}
Using the normalization of the state $\braket{\psi|\psi}=1$ to write $\braket{\dot{\psi}|\psi}=-\braket{\psi|\dot{\psi}}$, we can rewrite the expression as:
\al{
    \dot{\hat{\gamma}}^2 = -&\ket{\dot{\psi}}\!\!\braket{\dot{\psi}|\psi}\!\!\bra{\psi} + \ket{\dot{\psi}}\!\!\bra{\dot{\psi}}\notag\\
        +&\ket{\psi}\!\!\braket{\dot{\psi}|\dot{\psi}}\!\!\bra{\psi} - \ket{\psi}\!\!\braket{\psi|\dot{\psi}}\!\!\bra{\dot{\psi}}\,.
}
Hence, upon taking the trace, cyclicity and completeness allow the expression to take the following form:
\al{
    \Tr{\dot{\gamma}^2} = 2\bra{\dot\psi}\p{1 - \ketbra{\psi}{\psi}}\ket{\dot\psi}\,,
}
which is nothing more than a multiple of the Fubini-Study metric on the projective space in the ray space of physically distinct states \cite{Nakahara}. This fixes the metric since $\kappa=1/2$ is required to maintain equality with the Fubini-Study metric in the pure limit. In the end, we define
\al{
    g_{\mu\nu}^{(\text{1RDM})} \equiv \frac{1}{2}\Tr{\partial_{\mu}\op{\gamma}\partial_{\nu}\op{\gamma}}\ddp[\lambda]{x^{\mu}}\ddp[\lambda]{x^{\nu}}\,,\label{eq:metric}
}
such that
\al{
    \text{d}s^2 = g_{\mu\nu}\text{d}x^{\mu}\text{d}x^{\nu}\,,
}
as it should.


\bibliography{ref}